\documentclass[lettersize,journal]{IEEEtran}
\usepackage{amsmath,amsfonts}
\pdfoutput=1
\usepackage{algorithmic}
\usepackage{algorithm}
\usepackage{array}
\usepackage{textcomp}
\usepackage{stfloats}
\usepackage{url}
\usepackage{verbatim}
\usepackage{graphicx}
\usepackage{cite}
\usepackage{multirow,tabularx}
\usepackage{epsfig}
\usepackage{subfigure}
\usepackage{color,xcolor}
\usepackage{algorithmic}
\usepackage{amsmath}
\usepackage{amssymb,amsthm}
\usepackage{makecell}

\begin{document}

\title{A Survey on Multi-Behavior Sequential Recommendation}


\author{
Xiaoqing Chen,
Zhitao~Li,
Weike Pan, 
and~Zhong Ming
\IEEEcompsocitemizethanks{\IEEEcompsocthanksitem Xiaoqing Chen, Zhitao~Li,  Weike~Pan and Zhong~Ming are with the 
College of Computer Science and Software Engineering, Shenzhen University, Shenzhen, China, 518060.
}

}

\markboth{IEEE TRANSACTIONS ON KNOWLEDGE AND DATA ENGINEERING}%
{Shell \MakeLowercase{\textit{et al.}}: Bare Demo of IEEEtran.cls for Computer Society Journals}


\maketitle

\begin{abstract}
People usually have the explicit or implicit desire to get the information they need and are most interested in from \textcolor{black}{massive} information, which has led to the creation of personalized recommender systems. Recommender systems \textcolor{black}{are set up to address the issue of} information overload in traditional information retrieval systems such as search engines, and have been \textcolor{black}{a significant area of research focused on recommending} information that is of most interest to users. There is a sequential nature to the behavior of a person interacting with a system, such as examining one item of clothing before examining others. The problem of taking this sequential nature into account in delivering recommendation is known as sequential recommendation (SR). The traditional sequential recommendation problem \textcolor{black}{merely takes into account} a single type of behavior of the users, while in real scenarios users tend to engage in multiple types of behaviors, such as examining and adding clothes to cart before purchasing them, leading to the proposal of multi-behavior sequential recommendation (MBSR). MBSR considers both sequentiality and heterogeneity of user behaviors, which can achieve state-of-the-art recommendation \textcolor{black}{through} suitable modeling. Hence, MBSR is a relatively new and worthy direction for in-depth research, for which some related works have been proposed. This survey aims to shed light on the MBSR problem. Firstly, we introduce MBSR in detail, including its problem definition, application scenarios and challenges faced. Secondly, we detail the classification of MBSR, including neighborhood-based methods, matrix factorization-based methods and deep learning-based methods, where we further classify the deep learning-based methods into different learning architectures based on RNN, GNN, Transformer, and generic architectures as well as architectures that integrate hybrid techniques. In each method, we present related works based on the data perspective and the modeling perspective, as well as analyze the \textcolor{black}{strengths, weaknesses and features} of these works. Finally, we discuss some promising future research directions to address the challenges and improve the current status of MBSR.
\end{abstract}

\section{Introduction}\label{sec:introduction}
Nowadays, people are increasingly relying on the Internet to obtain information, and are faced with information overload due to the complexity and huge amount of network information. Traditional search engines cannot filter the items for each user well, making it difficult for people to quickly access the information they want. As such, recommender systems that can effectively solve the information overload problem and provide personalized services to different users are of great importance. Recommender systems~\cite{adomavicius2005recommender-system,Book2015-RecSysHandbook} is a fundamental tool to recommend items of most interest to the users from a large amount of information. The recommendation process usually involves collecting and analyzing the users' historical behavior data to learn their preferences and behavior patterns, and thus find the items which better align with their preferences. 
The historical behavior data used \textcolor{black}{can be divided into explicit feedback and implicit feedback. The explicit feedback data, also known as multi-class feedback, includes the behaviors such as a user’s ratings and likes on items, while the implicit feedback data, also known as one-class feedback, includes the behaviors like browses, adds to cart and purchases.}

Since the implicit feedback data is more readily available compared with the explicit feedback data in real scenarios, many works have focused on studying recommendation problems based on \textcolor{black}{one single type of implicit feedback behavior, which brings up the issue of single-behavior recommendation (SBR)~\cite{SPAC2014-DCTR, Neurocomputing2016-PPMF, UAI2009-BPR}. However, SBR usually contains fewer data, which is prone to the data sparsity or cold-start issues~\cite{KBS2015-ABPR}.} Besides, there is often more than one type of interaction between users and items in real-world scenarios, \textcolor{black}{such as examination, adding to cart and purchase in the setting of an e-commerce platform.} This indicates that there will be multiple types of user feedback, i.e., users' feedback is heterogeneous, which needs to be taken into account when modeling user preferences~\cite{KBS2015-ABPR, InformationSciences2018-BPRH, KDD2016-MultipleTypesFeedbackEmpiricalStudy}. As such, researchers have turned to the study of \textcolor{black}{multi-behavior recommendation (MBR) problems~\cite{IS2016-TranferLearning, TOIS2019-TransfertoRank}}. Different from the SBR problem, \textcolor{black}{MBR} provides personalized recommendations to users based on their heterogeneous one-class implicit feedback. In the modeling of MBR, not only a user's target feedback, such as the purchase behavior during online shopping, but also the information from auxiliary feedback such as browsing and favorites will be considered.

In addition to being heterogeneous, users' implicit feedback is also naturally sequential. In order to utilize the sequential information, some recent works have proposed single-behavior sequential recommendation (SBSR),
which begins to consider the sequential information of the one-class implicit feedback from users. Currently, there are advanced methods with better recommendation performance in SBSR. They use the sequences of users’ historical behaviors composed of one-class feedback to learn users' long-term and short-term preferences, and thus predict the items that users may interact with in the near future~\cite{WWW2010-FPMC, TOIS2019-BIS, WSDM2018-Caser}. However, SBSR only models the users' one-class implicit feedback, ignoring other implicit feedback information from users. Multi-behavior sequential recommendation (MBSR),
can fully consider the implicit feedback information of the users, solving the problem in SBSR. In MBSR, an algorithm usually models a user's heterogeneous behaviors and the sequential information of the behaviors simultaneously, so as to recommend items more aligned with the real preferences of the user~\cite{TKDE2017-RLBL, SIGIR2020-MKM-SR}. With the integration of MBR and SBSR, MBSR can be closer to the real behaviors of users. Nonetheless, it correspondingly brings more new challenges, including (i) sequence modeling of heterogeneous behavioral feedback, (ii) relationship modeling between user behaviors, (iii) joint long-term and short-term preferences modeling, (iv) existing noise, bias and other related issues in MBSR. We will discuss the specific challenges in detail in a subsequent section.

We also review some relatively well-known works and state-of-the-art works from leading conferences and journals in this survey, so as to illustrate how the MBSR problem is studied for personalized recommendation, hoping to provide some guidance for future research on MBSR. Specifically, we categorize the existing works on multi-behavior sequential recommendation into neighborhood-based methods, matrix factorization-based methods, and deep learning-based methods. Firstly, in neighborhood-based methods, we introduce how to use the neighborhood information to solve the recommendation problem, and how to extend existing methods from SBSR to MBSR. Secondly, in matrix factorization-based methods, we introduce the general idea of matrix factorization used in recommendation problems and mainly introduce the typical work from the perspective of transfer learning. Thirdly, in deep learning-based methods, we mainly focus on how to apply the ideas of deep learning to the MBSR problem, and describe the MBSR methods based on recurrent neural network (RNN), graph neural network (GNN), Transformer, generic methods and hybrid methods. We further delineate the related works of each deep learning framework from different modeling perspectives. Finally, we briefly discuss some future research directions, and conclude the paper.

The multi-behavioral sequential recommendation is currently considered an emerging area with limited related works and a lack of relevant surveys. In order to provide a comprehensive overview and enable researchers to keep abreast of the latest developments in MBSR, we conduct a survey on this topic, where we classify and compare various techniques and related works. The key contributions of our survey are summarized as follows.
\begin{itemize}
\item[$\bullet$]We present an in-depth overview of the MBSR problem by discussing its background, problem definition, application scenarios, and the existing challenges. Additionally, we provide a comprehensive classification of the current works on MBSR from three key perspectives: technique, data, and modeling.
\item[$\bullet$]We provide a summary of the strengths and weaknesses of each technique employed, alongside a detailed comparison and analysis of representative MBSR works based on the provided classification.
\item[$\bullet$]We propose valuable future research directions to address the challenges posed by MBSR.
\end{itemize}

The rest of the paper is organized as follows. In Section 2, we give the background on MBSR, \textcolor{black}{encompassing} four aspects, i.e., problem definition, application scenarios, challenges and categorization. Then in Sections 3, 4 and 5, we present an outline of the prominent works of MBSR in terms of neighborhood-based methods, matrix factorization-based methods and deep learning-based methods, respectively. In Section 6, we discuss some possible future research directions for MBSR, and finally, we draw the paper to a conclusion in Section 7.

\section{Preliminaries}
\subsection{Problem Definition}
\begin{figure}[!htb]
	
	\begin{center}
		
		\begin{tabular}{cc}
			
			\includegraphics[width=3.3in]{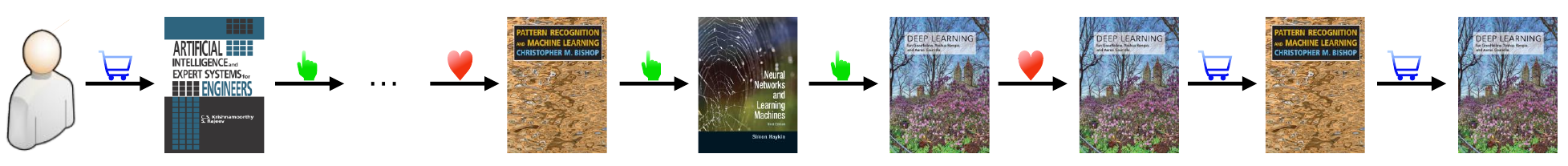}
			
		\end{tabular}
		
	\end{center}
	
	\caption{Illustration of multi-behavior sequential recommendation (MBSR).}
	
	\label{fig1:problem definition of MBSR}
\end{figure}

The MBSR problem mainly focuses on next item recommendation in a heterogeneous feedback sequence. We assume that there is a set of users, i.e., $\mathcal{U}$, a set of items, i.e., $\mathcal{I}$, and a set of behaviors (or feedback), i.e., $\mathcal{F}$, in the system. For the corresponding recommendation methods, the input is a set of (user, heterogeneous behavior sequence) pairs, i.e., $\left(u, \mathcal{S}_{u}\right)$, where $u \in \mathcal{U}$ represents a user ID, and $\mathcal{S}_{u}=\left\{\left(i_{u}^{1}, f_{u}^{1}\right), \ldots,\left(i_{u}^{t}, f_{u}^{t}\right), \ldots,\left(i_{u}^{\left|\mathcal{S}_{u}\right|}, f_{u}^{\left|\mathcal{S}_{u}\right|}\right)\right\}$ represents the historical interaction sequence between the user and the items. In the sequence, each $\left(i_{u}^{t}, f_{u}^{t}\right)$ tuple represents the (item, behavior) pair composed of the item $i$ and the corresponding behavior $f$ interacted by the user $u$ at the $t$th time step, where $i_{u}^{\textcolor{black}{t}} \in \mathcal{I}$, $\quad f_{u}^{\textcolor{black}{t}} \in \mathcal{F}$. We use time step instead of timestamp since MBSR does not generally introduce timestamp which represents precise time in modeling. Proper modeling the input data can learn the user’s preferences, as well as the representations and relationships of the items and behaviors, etc. Based on a typical recommendation method, we can predict the preference value $\hat{r}_{t+1,j}^{u}$ of a user $u$ for any item $j \in \mathcal{I}$ at the $(t+1)$th \textcolor{black}{time step} according to \textcolor{black}{the most recent $L$ historical interactions of user $u$ before the $(t+1)$th \textcolor{black}{time step},} i.e., $\mathcal{S}_{u}^{t}=\left\{\left(i_{u}^{t-L+1}, f_{u}^{t-L+1}\right), \ldots,\left(i_{u}^{\ell}, f_{u}^{\ell}\right), \ldots,\left(i_{u}^{t}, f_{u}^{t}\right)\right\}, t \leq \left|\mathcal{S}_{u}\right|$. We can then rank the preference values $\hat{r}_{\left|\mathcal{S}^{u}\right|+1,j}^{u}$ of the user $u$ for the candidate items $j \in \mathcal{I}$ to generate a top-$K$ list of items for user $u$, which indicates the next items that user $u$ is most likely to interact with.
We illustrate the general MBSR in Figure~\ref{fig1:problem definition of MBSR}, and show the commonly used symbols and corresponding interpretations in Table~\ref{tbl:notation}, where \textcolor{black}{we employ various font styles to denote diverse types of notations,} i.e., uppercase bold for matrices, lowercase bold for vectors, lowercase non-bold for scalars, and copperplate for sets.

\begin{table*}[htbp]
\caption{Some notations and explanations.} \label{tbl:notation}
\begin{center}
 \small
\begin{tabular}{ p{2.5cm} | p{12cm} } \hline\hline

Notation & Explanation\\
    \hline\hline

$n$
    & user number\\
$m$
    & item number\\
$\mathcal{R}$
    & the observed set of (user, item, behavior) tuples\\
$\mathcal{U}$
    & user set\\
$\mathcal{I}$
    & item set\\
$\mathcal{F}$
    & behavior set\\
$u \in \mathcal{U}$
    & user ID\\
$i \in \mathcal{I}$
    & item ID\\
$f \in \mathcal{F}$
    & behavior ID\\ 
    \hline
$\mathcal{S}_u$
    & the sequence of (item, behavior) pairs that user $u$ has interacted with\\
$\mathcal{S}_{\mathrm{e}}$
    & the sequence examined by user $u$\\
$\mathcal{S}_{\mathrm{n}}$
    & the sequence unexamined by user $u$\\
$\mathcal{S}_{\mathrm{l}}$
    & the sequence that user $u$ has liked\\
$\mathcal{S}_{\mathrm{d}}$
    & the sequence that user $u$ has disliked\\
    \hline
${i}_{u}^{t}\in \mathcal{I}$
    & the item interacted by user $u$ at \textcolor{black}{the time step} $t$ ($t \in \{1, 2, \ldots, \left| \mathcal{S}_u \right| \}$)\\
${f}_{u}^{t}\in \mathcal{F}$
    & the behavior of user $u$ at \textcolor{black}{the time step} $t$ ($t \in \{1, 2, \ldots, \left| \mathcal{S}_u \right| \}$)\\
$U_u \in \mathbb{R}^{d \times 1}$
    & the \textcolor{black}{embedding} of user $u$\\
$V_i \in \mathbb{R}^{d \times 1}$
    & the \textcolor{black}{embedding} of item $i$\\
$F_f \in \mathbb{R}^{d \times 1}$
    & the \textcolor{black}{embedding} of behavior $f$\\
    \hline
$\hat{r}_{t,\textcolor{black}{i}}^{u}$
    & the predicted preference of user $u$ to item $\textcolor{black}{i}$ at \textcolor{black}{the time step} $t$\\
$\hat{r}_{\textcolor{black}{t},i}$
    & the predicted preference on item $i$ at \textcolor{black}{the time step} $\textcolor{black}{t}$\\ 
$\hat{r}_{\textcolor{black}{t},i,f}^{u}$
    & the predicted preference that user $u$ generates behavior $f$ on item $i$ at \textcolor{black}{the time step} $\textcolor{black}{t}$\\
    \hline
$\mbox{Em}(\cdot)$
    & the ID-to-embedding function\\
$\textcolor{black}{\sigma(\cdot)}$
    & \textcolor{black}{the sigmoid function}\\
$\textcolor{black}{\odot}$
    & \textcolor{black}{the element-wise product function}\\
    
\hline\hline
\end{tabular}
\end{center}
\end{table*}
\subsection{Application Scenarios}
\textcolor{black}{In recommender systems, MBSR is a relatively new research hotspot,} attracting extensive attention from both academia and industry. In industry, the related works on MBSR are mainly applied to the click-through rate (CTR) prediction tasks. \textcolor{black}{Similar to sequential recommendation, CTR leverages historical (user, item) interaction information to generate a list of items for recommendation to the user at the next time step. However, in contrast to sequential recommendation, CTR ranks items by predicting a user's click-through rate on the items, which makes the CTR task require an input of item information, and thus the data processing methods and models used will often be different from those in sequential recommendation.} Currently, MBSR has been used in \textcolor{black}{many areas, varying from e-commerce~\cite{smith2017Amazon,wang2018Alibaba} and video recommendation~\cite{covington2016Youtube,gomez2015Netflix} to news recommendation~\cite{liu2010news,wu2020news}}. In e-commerce, researchers predict the items that \textcolor{black}{are most probable to be purchased by users} through the behavior sequential information of users' browses, adds-to-cart, favorites, and purchases~\cite{TKDE2017-RLBL, KDD2018-BINN,CIKM2020-DMT}. In video recommendation, users will generate behaviors like \textcolor{black}{examinations}, shares and others, where the sharing behavior can be used as the target behavior, and the \textcolor{black}{examination} behavior can be used as the auxiliary behavior~\cite{WWW2020-MGNN-Spred}. In news recommendation, users may interact with news by explicit feedback (e.g., dislike) and implicit feedback (e.g., browse), which makes it possible to incorporate these feedbacks to infer users' positive and negative interest preferences~\cite{WWW2022-FeedRec}.

To better learn the real preferences of users, some research works also take into account the dwelling time, category information, and other more fine-grained information in the MBSR problem~\cite{WSDM2018-RIB, WSDM2020-HUP, TOIS2021-IARS}. 
Moreover, some works on MBSR focus not only on better capturing user preferences, but also on designing relevant multi-behavior \textcolor{black}{sequential} recommendation algorithms in the context of user privacy protection issues. Importantly, MBSR methods also consider the \textcolor{black}{heterogeneous behavior information of users and the sequential information within or between the behaviors}, allowing for more useful information to be learned when modeling, which makes it closer to real recommendation scenarios. Hence, it is of great significance to design a recommendation algorithm for \textcolor{black}{multi-behavior sequential recommendation}.

\subsection{Challenges}
The MBSR problem involves modeling both multiple behaviors and behavioral sequences, contributing to the necessity to consider the existing problems of MBR and SBSR, as well as how to integrate these two kinds of information well. In particular, MBSR has to face the following challenges.

\begin{itemize}
\item[$\bullet$]Sequence modeling of heterogeneous behavioral feedback. In a traditional sequential recommendation problem~\cite{ICLR2016-GRU4Rec, TOIS2019-BIS}, researchers mostly consider only a single type of behavior, ignoring the potential and importance of other behaviors, especially in instances where the utilized data for the target behavior is sparse. It indicates that it is necessary to model the users' multiple heterogeneous behaviors in the sequential recommendation problem. 
\textcolor{black}{However, different from SBSR, the uncertainty of users' intention due to heterogeneous behaviors makes it more challenging to predict the user’s preference in MBSR.}
Hence, it is a key and challenging issue to model heterogeneous behaviors well in sequential recommendations without information loss.
\item[$\bullet$]Relationship modeling between user behaviors. In the MBSR problem, multiple behaviors of users are often related to each other~\cite{WSDM2018-RIB}. For example, in an e-commerce platform, users tend to examine an item, and check the reviews of the item before purchasing it, or purchase an item after examination and adding to cart other items of the same category. Different from MBR, which does not consider the sequential relationship of behaviors, MBSR takes the sequential nature of various behaviors into account. For example, MBR treats both cases the same for users who examine first and then purchase and for users who purchase and then examine, whereas MBSR considers the distinction between the two in modeling. According to the above issues mentioned, there are correlations and transitions among different behaviors within a user-item interaction sequence, which is a great challenge in modeling.
\item[$\bullet$]Joint long-term and short-term preference modeling with heterogeneous behaviors of users. Most of the traditional recommendation algorithms statically model the interaction information between users and items~\cite{KDD2009-CFTemporalDynamics, ICDM2015-MiningIndecisiveness, KDD2016-CKE}, which usually reveals users' long-term stable preferences. However, it ignores the dynamic changes of users' sequential behavior interacted with items. The dynamics of user preferences~\cite{AAAI2016-MTD} indicate the user's current short-term preferences, which can be revealed in the dynamically changing behavioral sequential information of the user. Since the user's interactive behavior information is a behavior sequence that naturally evolved over time, the sequential information can dynamically display the user's long-term stable preferences and short-term needs. How to take the sequential information into account is the main challenge of SBSR. 
\textcolor{black}{However, compared with SBSR, the behavioral heterogeneity of MBSR leads to an even greater challenge in modeling a user’s long-term and short-term preferences simultaneously. }
\item[$\bullet$]Related issues such as noise and bias. Some previous works regard \textcolor{black}{unexamined} behavior and missing behavior as implicit negative feedback of users, or simply ignore them~\cite{CIKM2010-CFExplicit&Implicit, RecSys2018-MultiTask, KDD2018-DEERS}. \textcolor{black}{Unexamination} does not always represent a negative user preference~\cite{SIGIR2016-eALS}, nor does an \textcolor{black}{examination} represent a positive user preference, where there may be mis-\textcolor{black}{examinations}. As for bias, since most current recommender systems tend to utilize the implicit feedback of users to make recommendation to users with the goal of more accurate item ranking, there may be a selection bias in implicit feedback data (e.g., a user may \textcolor{black}{examine} on an item simply because it is ranked highly). As such, the possible noise and bias in MBSR also deserve our attention.
\end{itemize}
\begin{figure}[!htb]
	\begin{center}
		\begin{tabular}{cc}
			\includegraphics[width=3.6in]{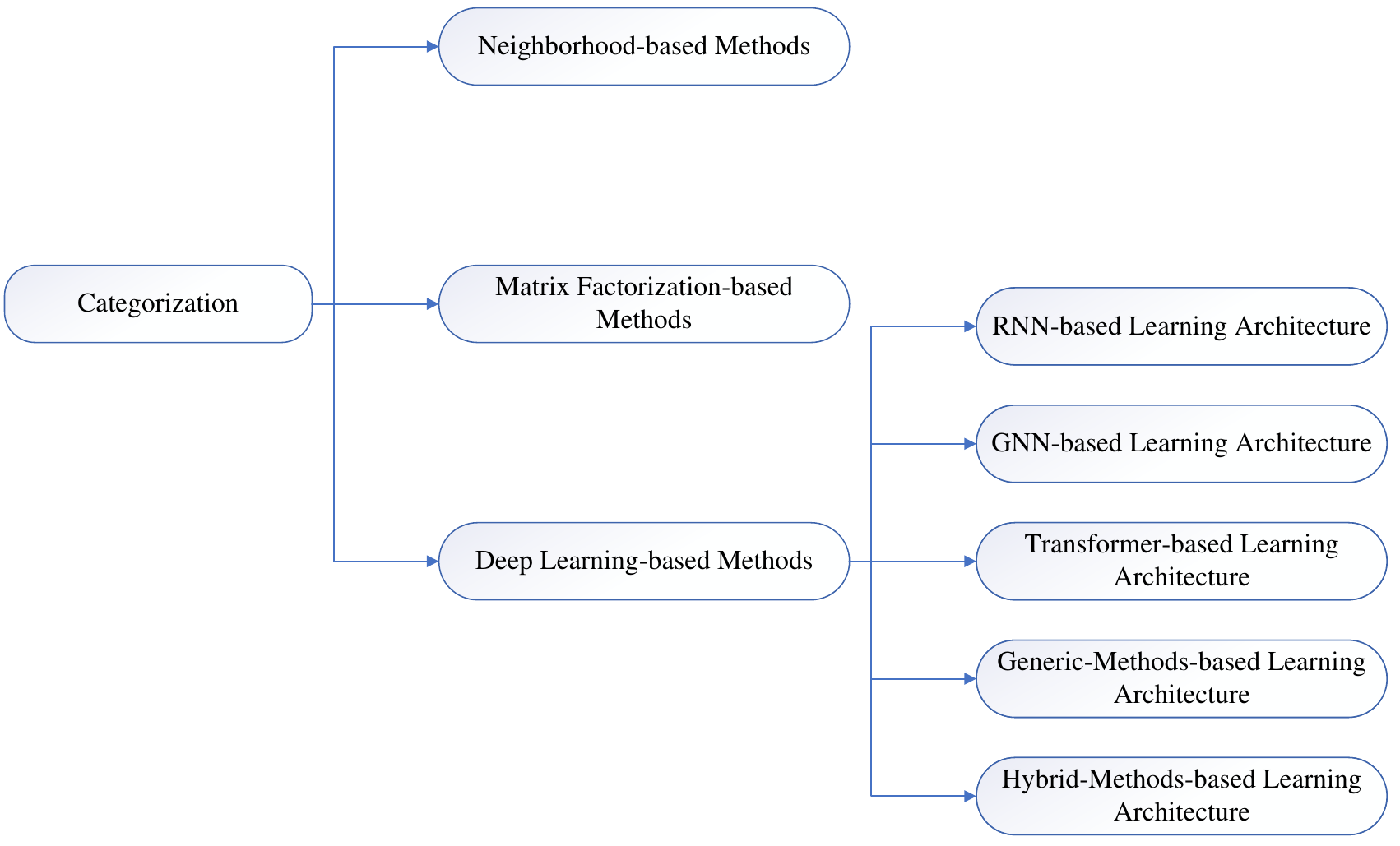}
		\end{tabular}
	\end{center}
	\caption{Categorization of multi-behavior sequential recommendation (MBSR).}
	\label{fig2:categorization}
\end{figure}
\subsection{Categorization}
The relevant methods applied to the MBSR problem can be classified into three categories at the technical level, including neighborhood-based methods, matrix factorization-based methods and deep learning-based methods, which are also the three major methods generally included in the field of recommender systems~\cite{Book2015-RecSysHandbook,TOIS2020-HOCCFsurvey}. For MBSR, \textcolor{black}{with} the continuous rise of artificial neural network and deep learning in various fields in recent years, and \textcolor{black}{with the increasing} complexity of information (i.e., the necessity to model heterogeneous behavior information and behavior sequential information simultaneously), as well as for higher recommendation performance, \textcolor{black}{the majority of the current works} utilize deep learning-based methods for modeling, with fewer efforts on the other two branches. As a result, we will \textcolor{black}{primarily pay attention to} the deep learning-based methods for the MBSR problem in this paper.

For deep learning-based methods, we review the classic, the best and the latest works according to the neural network architectures they utilize, including RNN-based learning architecture, GNN-based learning architecture, Transformer-based learning architecture, generic-methods-based learning architecture and hybrid-methods-based learning architecture. We first introduce the basic paradigm of each type of neural network architecture, and then discuss the related works applied to the MBSR problem. In these works, the loss functions used generally contain pointwise-based loss function, such as logistic loss~\cite{ANIPS2014-LogisticMF} and square loss~\cite{ANIPS2007-PMF} and pairwise-based loss function like BPR loss~\cite{UAI2009-BPR}, as well as their combined or varied loss functions. In the second place, we classify the related works according to different data perspectives and modeling perspectives under each neural network architecture. \textcolor{black}{Specifically, the data perspectives comprise four forms, namely a sequence of (item, behavior) pairs, some behavior-specific subsequences of items, a behavior-agnostic sequence of items and a sequence of behaviors; the modeling perspectives include both local and global approaches; and some of the MBSR works may integrate different data or modeling perspectives.}
As for related works with each neural network architecture, we discuss \textcolor{black}{the strengths, weaknesses, features} and some issues related to sequential heterogeneous information, such as considering more fine-grained information in modeling (e.g., \textcolor{black}{item category}), and taking into account how to address the noise and bias. We illustrate the above detailed categorization in Figure~\ref{fig2:categorization}.

\section{Neighborhood-based Methods}
The neighborhood-based method~\cite{desrosiers2011neighboorhood} is an early method in recommender systems, which is mainly divided into user-based collaborative filtering, item-based collaborative filtering and hybrid collaborative filtering.
\textcolor{black}{In the idea of user-based collaborative filtering, two users who have similar taste in the past may also have similar taste in the future, while in the idea of item-based collaborative filtering, users may purchase items similar to those purchased in the past. Hybrid collaborative filtering combines the ideas of the first two, and its prediction is a weighted combination of them. Regardless of the form of neighborhood-based methods, the main core concept of which is similarity.}
However, \textcolor{black}{how to define the similarity is a matter of concern for MBSR.}
\subsection{Basic Paradigm}
In the case with implicit feedback data, the similarity between two items can be calculated using measures such as \textcolor{black}{the} Jaccard index and \textcolor{black}{the} cosine similarity.
\textcolor{black}{Taking item-based collaborative filtering with the Jaccard index as an example}, the similarity of items $k$ and $j$ is calculated as follows:
\begin{align}
    s_{\textcolor{black}{i i^{\prime}}} = \frac{\left|\mathcal{U}_{\textcolor{black}{i}} \cap \mathcal{U}_{\textcolor{black}{i^{\prime}}}\right|}{\left|\mathcal{U}_{\textcolor{black}{i}} \cup \mathcal{U}_{\textcolor{black}{i^{\prime}}}\right|}
\end{align}
\textcolor{black}{where $\mathcal{U}_{\textcolor{black}{i}}$ and $\mathcal{U}_{\textcolor{black}{i^{\prime}}}$ denote the set of users who have interacted with item $i$ and item $i^{\prime}$, respectively.}

Based on the calculated similarity, we may select the top-$K$ nearest item set $\mathcal{N}_{\textcolor{black}{i^{\prime}}}$ for each item $i^{\prime}$, and then predict the score according to the following formula:
\begin{align}
    \hat{r}_{u \textcolor{black}{i^{\prime}}}=\sum_{{\textcolor{black}{i}} \in \mathcal{I}_u\cap  \mathcal{N}_{\textcolor{black}{i^{\prime}}}} s_{\textcolor{black}{i i^{\prime}}}
\end{align}

As the predicted rating of an item increases, the possibility that the user will be interested in it increases accordingly. \textcolor{black}{Although there is almost no work for MBSR using neighborhood-based methods, we will introduce BIS~\cite{TOIS2019-BIS}, a work toward the SBSR problem, to illustrate the idea of the use of similarity in sequential recommendation. It is expected to have some possibilities and inspirations to solve the MBSR problem.}

\subsection{BIS}
\textcolor{black}{Bidirectional item similarity (BIS) designs a bidirectional item similarity to perform the next-item recommendation task.}
The bidirectional item similarity between items $i$ and $i^{\prime}$ is defined as follows:
\begin{align}
	&\mbox{sim}_{i^{\prime} \rightarrow i}^{(\ell, \rho)}=\frac{\sum_{u \in \textcolor{black}{\mathcal{U}_{i} \cap \mathcal{U}_{i^{\prime}}}} \delta\left(-\rho \ell \leqslant\left(t-t^{\prime}\right) \leqslant \ell\right)} {\left|\mathcal{U}_{i} \cup \mathcal{U}_{i^{\prime}}\right|}
\end{align}
where $\ell$ and $\rho$ are hyperparameters. \textcolor{black}{In this equation, if the condition $-\rho \ell \leqslant\left(t-t^{\prime}\right) \leqslant \ell$ is satisfied, $\delta\left(-\rho \ell \leqslant\left(t-t^{\prime}\right) \leqslant \ell\right)$ will be set to 1, so that the numerator will be added by 1 and the similarity between items $i$ and $i^{\prime}$ will increase accordingly.} It is worth noting that when $\rho$ is equal to 1 and $\ell \rightarrow \infty$, the bidirectional item similarity degenerates to Jaccard index, i.e.,$\mbox{sim}_{i^{\prime} \rightarrow i}^{(\infty, 1)}=\frac{\left|\mathcal{U}_{i} \cap \mathcal{U}_{i^{\prime}}\right|}{\left|\mathcal{U}_{i} \cup \mathcal{U}_{i^{\prime}}\right|}$, which does not take into account any sequential information.
\textcolor{black}{When predicting the preference score, BIS only considers the bidirectional item similarities of the last $k$ items that user $u$ has interacted with.}

\textcolor{black}{Obviously, BIS and ABIS (adaptive BIS)~\cite{TOIS2019-BIS}, an improved version of BIS based on some factorization techniques, can be extended to the problem of multi-behavior sequential recommendation. For example, if we divide the input sequence into multiple behavior-specific subsequences, we can easily apply BIS and ABIS for each subsequence. These neighborhood-based methods are easy to maintain and more interpretable, but they are less able to capture user preferences and lack transitivity,} which means that two users will never be connected if they have not bought a common item. Moreover, ABIS only considers the closest neighboring items in modeling, ignoring users' long-term preferences and periodicity.

\section{Matrix Factorization-based Methods}
\textcolor{black}{Although neighborhood-based methods may provide interpretability, their aforementioned disadvantages and lower efficiency make them less applicable to MBSR.} To address the problem of non-transitivity, \textcolor{black}{a method named matrix factorization has been proposed to connect} users who have not purchased common items before~\cite{ma2008mf,mehta2017mf}.
\subsection{Basic Paradigm}
In recommender systems, the idea of matrix factorization is mainly reflected in transforming the (user, item) interaction matrix into the inner product of two low-rank matrices, i.e., a user-specific matrix and an item-specific matrix. Taking the rating matrix \textcolor{black}{$\boldsymbol{M} \in \mathbb{R}^{m \times n}$} formed by (user, item) interactions as an example, $\boldsymbol{M}$ is decomposed into a user matrix \textcolor{black}{$\boldsymbol{U} \in \mathbb{R}^{m \times k}$} and an item matrix \textcolor{black}{$\boldsymbol{V} \in \mathbb{R}^{n \times k}$}, so that each missing value \textcolor{black}{(i.e., a predicted value)} $\hat{r}_{ui}$ in the rating matrix can be obtained by multiplying the user \textcolor{black}{embedding} $U_u$ and the item \textcolor{black}{embedding} $V_i$:
\begin{align}
    \hat{r}_{ui}=U_{u} \cdot V_{i}^{\top}
\end{align}

\subsection{TransRec++}
\begin{figure}[!htb]
	
	\begin{center}
		
		\begin{tabular}{cc}
			
			\includegraphics[width=2.5in]{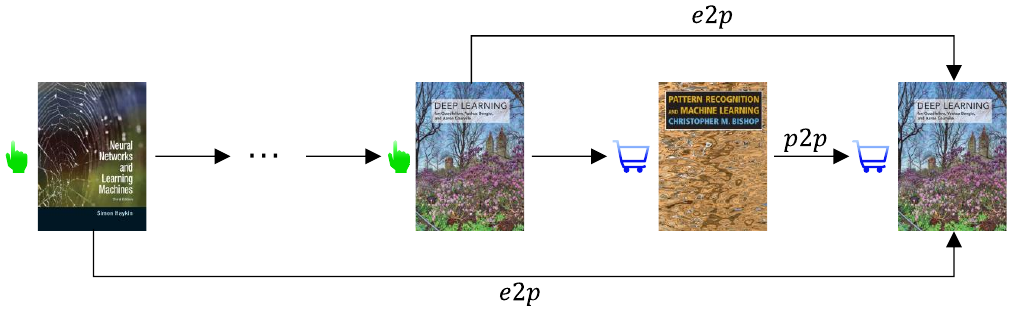}
			
		\end{tabular}
		
	\end{center}
	
	\caption{Illustration of behavior transition.}
	
	\label{fig:behavior transition}
\end{figure}

TransRec++~\cite{CS2022-TransRec++} introduces \textcolor{black}{several behavior transition vectors to capture the sequential relationships between user behaviors and their dynamics, and takes into account some recent preceding items which can learn the weights automatically.} The behavior transition vectors contain four types, i.e., from examination to examination $e2e$, from examination to purchase $e2p$, from purchase to examination $p2e$, and from purchase to purchase $p2p$, which we illustrate in Figure~\ref{fig:behavior transition}. 

In step $\ell$, the overall translation vector of user $u$ to the target item $i_{u}^{t}$ is calculated \textcolor{black}{by the following equation:}
\begin{align}
	&\tilde{U}_{u}^{(\ell) i_{u}^{t}}=U_{u}+U_{u}^{(\ell) b(i_{u}^{t-\ell}) 2 b(i_{u}^{t})}
\end{align}
where \textcolor{black}{$b(\cdot)$ denotes the behavior type.}
To achieve a transition of item $i_{u}^{t-\ell}$ to a future item $i_{u}^{t}$ in step $\ell$ for a user’s sequence, the formula can be calculated as follows:
\begin{align}
	&V_{i_{u}^{t-\ell}}+\tilde{U}_{u}^{(\ell) i_{u}^{t}} \approx V_{i_{u}^{t}}, \ell=1,2, \ldots, L
\end{align}
where $V_{i_{u}^{t-\ell}}$ and $V_{i_{u}^{t}}$ are the embedding vectors of item $i_{u}^{t-\ell}$ and item $i_{u}^{t}$, respectively.
The prediction formula is defined as follows:
\textcolor{black}{
\begin{align}
	&\hat{r}_{u{i_u^t}}=p_{i_{u}^t}-\sum_{\ell=1}^{L}\left(\eta_{\ell}+\eta_{\ell}^{u}\right) \mid\mid V_{i_{u}^{t-\ell}}+\tilde{U}_{u}^{(\ell) i_{u}^{t}}-V_{i_{u}^t}\|_{2}^{2}
\end{align}}where $p_{i_{u}}$ is an item bias, and $\eta_{\ell}$, $\eta_{\ell}^{u}$ denote \textcolor{black}{a global weight and a user-specific weight}, respectively.

\textcolor{black}{As one of the few matrix factorization-based solutions towards the next-item recommendation in MBSR, TransRec++} combines the ideas of Fossil~\cite{ICDM2016-Fossil} and TransRec~\cite{RecSys2017-TransRec} to address behavioral heterogeneity well.
\textcolor{black}{The proposed behavior transition can also be utilized in other deep learning-based approaches to reach better performance, such as RIB~\cite{WSDM2018-RIB} and BINN~\cite{KDD2018-BINN} that we will mention later.}
However, TransRec++ becomes more complex in modeling when there are more behavior types\textcolor{black}{, and when it only contains two types of behaviors its time complexity is already five times that of the SBSR-oriented method TransRec. }There may also be noise in the modeling of behavior transition, e.g., e2e may be caused by user mis-examination.

\textcolor{black}{In summary, as a conventional approach, the matrix factorization-based recommendation algorithm owns several benefits}, including high interpretability and computational efficiency. These algorithms employ a linear model, which possesses a straightforward structure, and a clear association between the modeling concept and the problem under consideration, leading to a higher degree of interpretability. The algorithms are computationally efficient as the model has few parameters, and typically, only matrix multiplication operations are necessary for computation. However, matrix factorization-based recommendation algorithms face challenges in handling non-linear features like sequence information and higher-order neighborhood information.

\section{Deep Learning-based Methods}
Due to the insignificant improvement in recommendation effectiveness of matrix factorization-based methods, researchers have turned to studying deep learning-based algorithms. Deep learning~\cite{lecun2015deepLearning,goodfellow2016deeplearning} is \textcolor{black}{an improvement on the traditional neural network}, and a multi-layer perceptron (MLP) with multiple hidden layers is a typical deep learning architecture. \textcolor{black}{Deep learning has made substantial progress in a variety of application areas, including} natural language processing and generation~\cite{ICML2008-MultitaskLearning, EMNLP2013-SemanticCompositionality}, speech recognition and synthesis~\cite{TASLP2011-CD-DNN-HMMs}, as well as computer vision~\cite{ANIPS2012-CNNforImagenet, ICASSP2013-FeatureDetector}. \textcolor{black}{Recently, deep learning has gained increasing usage} in recommender systems, demonstrating high recommendation performance in multi-behavior recommendation~\cite{TKDE2019-MultiCascadingBehaviors, IJCAI2019-MRMN}, sequential recommendation~\cite{ICLR2016-GRU4Rec, WSDM2018-Caser} and federated recommendation~\cite{arXiv2021-FedGNN,TIST2022-FedCTR}. In this section, we mainly \textcolor{black}{discuss} the application and the corresponding works of deep learning in MBSR in terms of different neural network architectures (i.e., RNN, GNN, Transformer, generic methods and hybrid methods).
\textcolor{black}{We will present how different works model sequentiality and heterogeneity in MBSR, and examine whether these works address some other specific challenges. Additionally, we will distinguish between the different applications of these works such as next-item, next-basket and session-based recommendation, and discuss their features, strengths, and weaknesses. This will help to establish a better understanding of the use of deep learning techniques in MBSR.}
\subsection{RNN-based Learning Architecture}
\subsubsection{Basic Paradigm}
Recurrent neural network (RNN)~\cite{mikolov2010RNN} is a classical deep learning method that can effectively process data with sequentiality. Currently, RNN has been applied to numerous fields, including information retrieval, speech recognition, machine translation and so on. Since RNN can take into account the characteristics of sequences, they have also been utilized to solve the SBSR problems and MBSR problems in the early works~\cite{ICLR2016-GRU4Rec, TKDE2017-RLBL, WSDM2018-RIB}.
\begin{figure}[!htb]
	
	\begin{center}
		
		\begin{tabular}{cc}
			
			\includegraphics[width=3.2in]{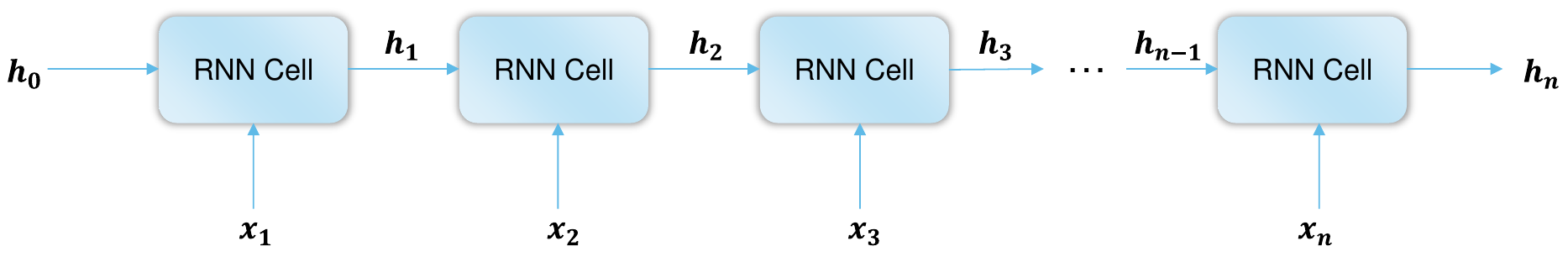}
			
		\end{tabular}
		
	\end{center}
	
	\caption{Illustration of recurrent neural network (RNN).}
	
	\label{fig:RNN}
\end{figure}

RNN contains multiple RNN cells, \textcolor{black}{with its basic structure illustrated in Figure}~\ref{fig:RNN}. In the RNN learning architecture, the current time step receives the output of the previous time step as the input, and the output obtained from the RNN cell will be used as the input of the next time step, so as to capture the sequential nature of the data. Each cell of RNN is a layer of deep feedforward neural network, and a set of learning parameters will be shared \textcolor{black}{across different time steps to capture sequential features and reduce the model complexity.} The basic formula of RNN is as follows:
\begin{align}
    &\boldsymbol{h}_{t}=\sigma\left(\boldsymbol{W}_{\textcolor{black}{\mathrm{xh}}} \boldsymbol{x}_{t}+\boldsymbol{W}_{\textcolor{black}{\mathrm{hh}}} \boldsymbol{h}_{t-1}+\boldsymbol{b}_{\textcolor{black}{\mathrm{h}}}\right) \\
    &\hat{\boldsymbol{y}}_{t}=\sigma\left(\boldsymbol{W}_{\textcolor{black}{\mathrm{ho}}} \boldsymbol{h}_{t}+\boldsymbol{b}_{\textcolor{black}{\mathrm{o}}}\right)
\end{align}
where $\boldsymbol{W}_{\textcolor{black}{\mathrm{xh}}} \textcolor{black}{\in \mathbb{R}^{d \times d}}$, $\boldsymbol{W}_{\textcolor{black}{\mathrm{hh}}} \textcolor{black}{\in \mathbb{R}^{d \times d}}$ and $\boldsymbol{W}_{\textcolor{black}{\mathrm{ho}}} \textcolor{black}{\in \mathbb{R}^{d \times d}}$ are the corresponding weight matrices, and $\boldsymbol{b}_{\textcolor{black}{\mathrm{h}}}$ and $\boldsymbol{b}_{\textcolor{black}{\mathrm{o}}}$ are the corresponding bias vectors.

However, there is a certain challenge in the training process of RNN, that is, as the depth deepens, RNN has the problem of gradient disappearance or gradient explosion, and thus is prone to the difficulty in dealing with the long-term dependency of data~\cite{bengio1994gradient}. To address this issue, many derivative methods based on RNN have been proposed, among which the most well-known ones are long short-term memory (LSTM)~\cite{hochreiter1997lstm} and gated recurrent unit (GRU)~\cite{chung2014GRU}, a simplified version of LSTM. Both LSTM and GRU set up a hidden unit in the hidden layer to store long-term features, which enables to address the issue of modeling long-term data dependency.

\subsubsection{Methods in MBSR}
There are some research works on RNN-based neural network architectures for solving MBSR problems, which differ in terms of the perspective of the input sequences and the perspective of modeling the behavior types. Specifically, from the data perspective, most of the works have an input sequence of (item, behavior) pairs, such as RLBL~\cite{TKDE2017-RLBL}, RIB~\cite{WSDM2018-RIB}, BINN~\cite{KDD2018-BINN}, HUP~\cite{WSDM2020-HUP}, IARS~\cite{TOIS2021-IARS} and DeepRec~\cite{WWW2021-DeepRec}. In contrast, other works have some behavior-specific subsequences of items, such as CBS~\cite{IJCAI2018-CBS}, DIPN~\cite{KDD2019-DIPN} and MBN~\cite{TKDD2022-MBN}. From the modeling perspective, \textcolor{black}{DeepRec models a user's behavior types in the cloud from a global perspective and in the user's own client from a local perspective, while other works mentioned above utilize a local perspective only to model the behavior types}. We distinguish and summarize these works in Table~\ref{tbl:RNN} and describe \textcolor{black}{some of them} in detail as shown below.
\begin{table*}[htbp]
\caption{Data \& modeling perspectives and features used in works based on RNN learning architecture.} \label{tbl:RNN}
\begin{center}
\small
\begin{tabular}{ p{1.5cm} | p{3cm} | p{2.5cm} |  p{9cm}} \hline\hline
Works & Data Perspective & Model Perspective & Features\\
    \hline\hline
\small
RLBL~\cite{TKDE2017-RLBL}
    & A sequence of (item, behavior) pairs
    & Local
    & \textcolor{black}{Capture the influence of heterogeneous behaviors by utilizing a behavior transition matrix.}\\
    \hline
RIB~\cite{WSDM2018-RIB}
    & A sequence of (item, behavior) pairs
    & Local
    & Leverage GRU and attention mechanism simultaneously.\\
    \hline
BINN~\cite{KDD2018-BINN}
    & A sequence of (item, behavior) pairs
    & Local
    & Design the CLSTM and the Bi-CLSTM, where the behavior vector is as context in LSTM.\\
    \hline
CBS~\cite{IJCAI2018-CBS}
    & Some behavior-specific subsequences of items
    & Local
    & Design of models with and without shared parameters for behaviors simultaneously; towards the next-basket recommendation.\\
    \hline
DIPN~\cite{KDD2019-DIPN}
    & Some behavior-specific subsequences of items
    & Local
    & Leverage GRU and attention mechanism simultaneously; behaviors are specific, including swipe, touch and browse interactive behavior.\\
    \hline
HUP~\cite{WSDM2020-HUP}
    & A sequence of (item, behavior) pairs
    & Local
    & Design the Behavior-LSTM where adds behavior gate and time gate to the LSTM; leverage attention mechanism; take into account the category of the items.\\
    \hline
IARS~\cite{TOIS2021-IARS}
    & A sequence of (item, behavior) pairs
    & Local
    & Propose Soft-MGRU (a multi-behavior gated recurrent unit) with sharing parameters between behaviors; leverage attention mechanism; take into account the category of the items.\\
    \hline
DeepRec~\cite{WWW2021-DeepRec}
    & \textcolor{black}{Some behavior-specific subsequences of items}
    & Local + Global
    & Utilizing multi-behavior sequence data to make privacy-preserving recommendation.\\
    \hline
MBN~\cite{TKDD2022-MBN}
    & Some behavior-specific subsequences of items
    & Local
    & The \textcolor{black}{overall} Meta-RNN and the \textcolor{black}{separate} Behavior-RNN share the learned potential representations by gathering and then scattering; towards the next-basket recommendation.\\
\hline\hline

\end{tabular}

\end{center}
\end{table*}

\begin{figure}[!htb]
	
	\begin{center}
		
		\begin{tabular}{cc}
			
			\includegraphics[width=2.5in]{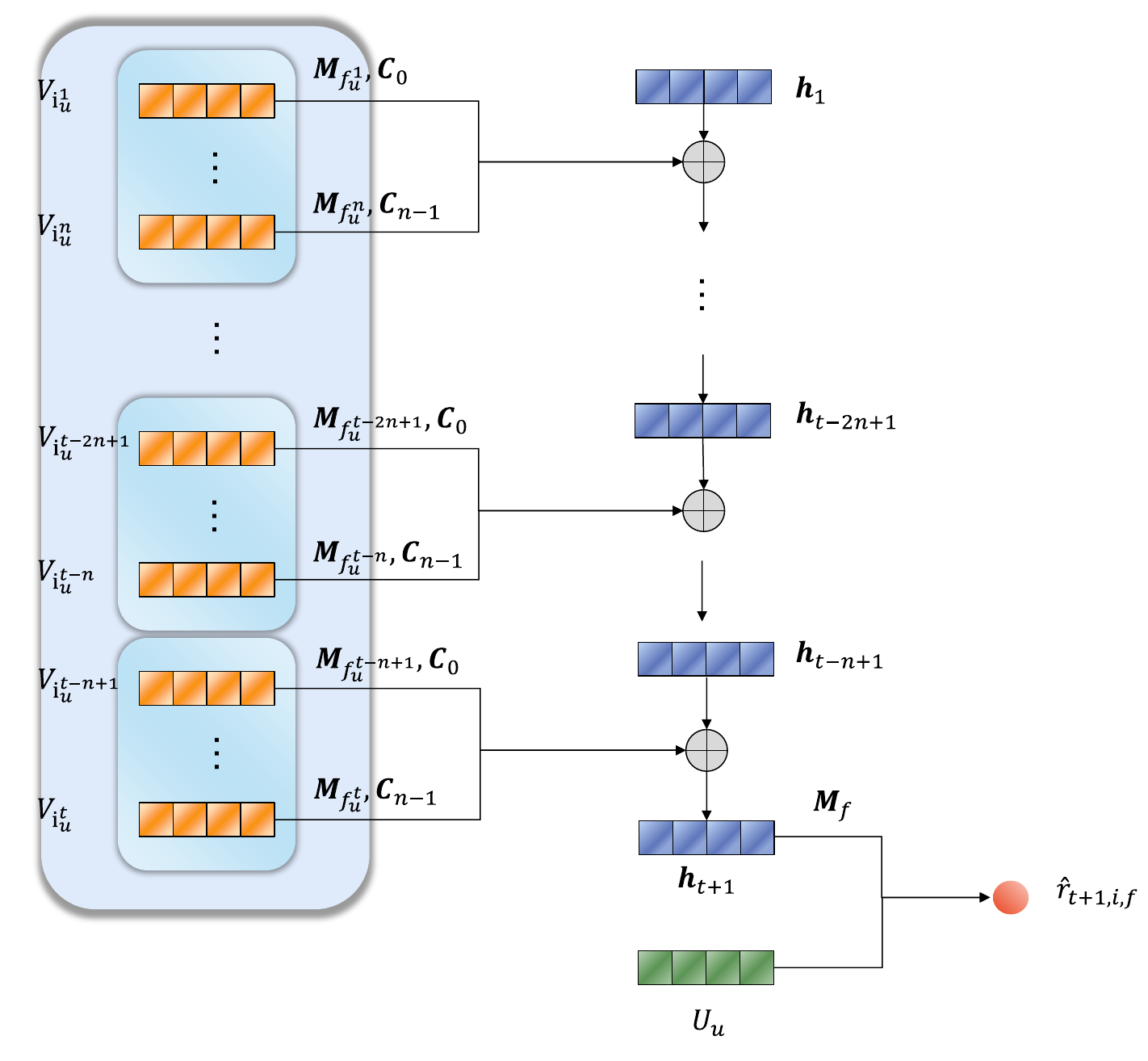}
			
		\end{tabular}
		
	\end{center}
	
	\caption{Illustration of recurrent log-bilinear model (RLBL).}
	
	\label{fig:RLBL}
\end{figure}

\textbf{RLBL.}
The recurrent log-bilinear model (RLBL)~\cite{TKDE2017-RLBL} \textcolor{black}{illustrated in Figure~\ref{fig:RLBL}} is the first work oriented towards \textcolor{black}{next-item recommendation}. RLBL integrates the ideas of RNN and log-bilinear (LBL) \textcolor{black}{to address the challenge of long-term and short-term preference modeling}. Specifically, RLBL uses behavior-specific transition matrices to \textcolor{black}{distinguish between heterogeneous behaviors} in a user's historical interaction sequence, and splits the sequence into multiple windows. Then RLBL captures the short-term contextual information for each window by LBL, and finally integrates these features at the granularity of the window by RNN to construct the user's long-term contextual information.

In RLBL, each window contains a sequence of (item, behavior) pairs of length $n$, i.e., $\{(i_{u}^{t-n+1},f_{u}^{t-n+1}),$ $..., \left(i_{u}^{t}, f_{u}^{t}\right)\textcolor{black}{\}}$. In the pair $\left(i_{u}^{t-i}, f_{u}^{t-i}\right)$ of the sequence, RLBL uses \textcolor{black}{an item embedding} $V_{i_{u}^{t-i}} \in \mathbb{R}^{d \times 1}$ to represent the historically interacted item $i_{u}^{t-i}$ of user $u$, \textcolor{black}{a behavior correlation embedding} $\boldsymbol{M}_{f_{u}^{t-i}} \in \mathbb{R}^{d \times d}$ to represent the user’s behavior $f_{u}^{t-i}$ for item $i_{u}^{t-i}$, and a position transition embedding $\boldsymbol{C}_{i} \in \mathbb{R}^{d \times d}$ to \textcolor{black}{separately} capture the position context information of each position in the window $\left(i_{u}^{t-i}, f_{u}^{t-i}\right), i \in\{0,1, \ldots, n-1\}$. Hence, the hidden state $\boldsymbol{h}_{t+1}$ at the $(t+1)$th time step is \textcolor{black}{calculated below:}
\begin{align}
    &\boldsymbol{h}_{t+1}=\boldsymbol{W}_{\mathrm{RLBL}} \boldsymbol{h}_{t-n+1}+\sum_{i=0}^{n-1} \boldsymbol{C}_{i} \boldsymbol{M}_{f_{u}^{t-i}} V_{i_{u}^{t-i}}
\end{align}
where $\boldsymbol{W}_{\text{RLBL}} \in \mathbb{R}^{d \times d}$ is utilized to capture the sequential information between the hidden state $\boldsymbol{h}_{t+1} \in \mathbb{R}^{d \times 1}$ and the hidden state $\boldsymbol{h}_{t-n+1} \in \mathbb{R}^{d \times 1}$. And then the predicted preference that user $u$ generates behavior $f$ on item $i$ at the $(t+1)$th \textcolor{black}{time step} is \textcolor{black}{calculated} as follows:
\begin{align}
    &\hat{r}_{t+1, i, f}=\left(\boldsymbol{h}_{t+1}+U_{u}\right)^{\mathrm{T}} \boldsymbol{M}_{f} V_{i}
\end{align}
where \textcolor{black}{$U_{u} \in \mathbb{R}^{d \times 1}$ is the user embedding, and} $\boldsymbol{h}_{t+1} \in \mathbb{R}^{d \times 1}$ is the \textcolor{black}{representation} incorporating the long-term and short-term preferences of user $u$.

\textcolor{black}{RLBL and its extended version TA-RLBL~\cite{TKDE2017-RLBL}, which considers continuous time differences, can model the long-term and short-term context information for data well with the consideration of the sequential and heterogeneous nature of user behaviors. However, the modeling of user behavior is relatively straightforward, and there are some important issues that are overlooked. For example, the transition matrix is the same for all users, and it does not take into account the feature information of the items.}

%
\textbf{RIB.}
An interpretable recommendation framework from the micro behavior perspective (RIB)~\cite{WSDM2018-RIB}, another classic work towards \textcolor{black}{next-item recommendation}, models heterogeneous behaviors and dwell time to capture more fine-grained user information \textcolor{black}{by GRU}. Specifically, RIB takes a sequence of (item, behavior) pairs as input, taking items and behaviors encoded as item embeddings and behavior embeddings via an embedding layer, respectively. Then the embedding $\boldsymbol{e}_{t} \in \mathbb{R}^{2 d \times 1}$ is obtained by concatenating the above two embeddings and \textcolor{black}{fed into a GRU layer to obtain the hidden state at each time step}. The calculation equations of the reset gate $\boldsymbol{r}_t \in \mathbb{R}^{d \times 1}$, the update gate $\boldsymbol{z}_t \in \mathbb{R}^{d \times 1}$, the internal state $\boldsymbol{c}_t \in \mathbb{R}^{d \times 1}$ and the external state $\boldsymbol{h}_t \in \mathbb{R}^{d \times 1}$ at the $t$th time step in GRU are \textcolor{black}{shown below:}
\begin{align}
    &\boldsymbol{r}_{t}=\sigma\left(\boldsymbol{W}_{\mathrm{er}} \boldsymbol{e}_{t}+\boldsymbol{W}_{\mathrm{hr}} \boldsymbol{h}_{t-1}\right)\\
    &\boldsymbol{z}_{t}=\sigma\left(\boldsymbol{W}_{\mathrm{ez}} \boldsymbol{e}_{t}+\boldsymbol{W}_{\mathrm{hz}} \boldsymbol{h}_{t-1}\right)\\
    &\boldsymbol{c}_{t}=\mathrm{tanh}\left(\boldsymbol{W}_{\mathrm{ec}} \boldsymbol{e}_{t}+\boldsymbol{W}_{\mathrm{hc}}\left(\boldsymbol{r}_{t} \cdot \boldsymbol{h}_{t-1}\right)\right)\\
    &\boldsymbol{h}_{t}=\left(1-\boldsymbol{z}_{t}\right) \boldsymbol{h}_{t-1}+\boldsymbol{z}_{t}\boldsymbol{c}_{t}
\end{align}
where $\boldsymbol{W}_{\mathrm{er}}$, $\boldsymbol{W}_{\mathrm{ez}}$, $\boldsymbol{W}_{\mathrm{ec}} \in \mathbb{R}^{d \times 2d}$, and $\boldsymbol{W}_{\mathrm{hr}}$, $\boldsymbol{W}_{\mathrm{hz}}$, $\boldsymbol{W}_{\mathrm{hc}} \in \mathbb{R}^{d \times d}$ are the \textcolor{black}{learnable weight parameters inside GRU.}
Then the hidden state is passed into an attention layer to get the attention score for each time step. Finally, in the output layer, the hidden states of each time step are multiplied with the corresponding attention scores, where the results are added to obtain a latent representation \textcolor{black}{for predicting the user's preference value for an item.}

\textcolor{black}{Similar to RLBL, RIB introduces different behavioral information into the input side of the RNN (in this case GRU), but the difference is that RIB uses an embedding matrix to represent multiple behavioral information, where each behavior corresponds to an embedding vector. RIB also uses an attention layer to capture the importance of different behaviors, and in the original paper, modeling of dwell time was also considered. Nevertheless, RIB may not capture real user behavior information since it uses an embedding matrix to represent behavior types and then concatenates them directly with the item embedding.}

	
		
			
			
		
	
	

\textbf{BINN.}
Behavior-intensive neural network (BINN)~\cite{KDD2018-BINN}, based on LSTM, models users' long-term and short-term preferences \textcolor{black}{to improve the next-item recommendation performance.}
BINN takes a sequence of (item, behavior) pairs as input and models each sequence from a local perspective. BINN contains two modules, session behaviors learning (SBL) \textcolor{black}{to model a user's current consumption motivation and preference behaviors learning (PBL) to learn the user's historical stable preference.} In SBL, a context-aware LSTM (CLSTM) incorporating the behavioral information as input is built,
\textcolor{black}{whose input gate $\boldsymbol{i}_t$, forgetting gate $\boldsymbol{f}_t$, output gate $\boldsymbol{o}_t $, internal state $\boldsymbol{c}_t$ and external state $\boldsymbol{h}_t$ at the $t$th time step are as follows:}
\begin{align}
    & \boldsymbol{i}_t = \sigma(\boldsymbol{W}_{\mathrm{vi}}V_{i_u^t}+\boldsymbol{W}_{\mathrm{hi}}\boldsymbol{h}_{t-1}+\boldsymbol{W}_{\mathrm{ci}} \boldsymbol{c}_{t-1}  +\boldsymbol{W}_{\mathrm{bi}} F_{f_u^t}+\boldsymbol{b}_{\mathrm{i}}) \\
&\boldsymbol{f}_t=\sigma(\boldsymbol{W}_{\mathrm{vf}} V_{i_u^t}+\boldsymbol{W}_{\mathrm{hf}}\boldsymbol{h}_{t-1}+\boldsymbol{W}_{\mathrm{cf}} \boldsymbol{c}_{t-1}+\boldsymbol{W}_{\mathrm{bf}} F_{f_u^t}+\boldsymbol{b}_{\mathrm{f}}) \\
    & \boldsymbol{c}_t=\boldsymbol{f}_t\boldsymbol{c}_{t-1}+\boldsymbol{i}_t\,\mathrm{tanh}(\boldsymbol{W}_{\mathrm{vc}}V_{i_u^t}+\boldsymbol{W}_{\mathrm{hc}}\boldsymbol{h}_{t-1} +\boldsymbol{W}_{\mathrm{bc}}F_{f_u^t}+\boldsymbol{b}_\mathrm{c}) \\
    & \boldsymbol{o}_t=\sigma(\boldsymbol{W}_{\mathrm{vo}}V_{i_u^t}+\boldsymbol{W}_{\mathrm{ho}}\boldsymbol{h}_{t-1}+\boldsymbol{W}_{mathrm{co}}\boldsymbol{c}_{t}+\boldsymbol{W}_{\mathrm{bo}}F_{f_u^t}+\boldsymbol{b}_\mathrm{o}) \\
    & \boldsymbol{h}_t= \boldsymbol{o}_t\,\mathrm{tanh}(\boldsymbol{c}_t)
\end{align}
where $\boldsymbol{W}_{(\cdot)} \in \mathbb{R}^{d \times d}$ are the internal model parameters of the LSTM. Then the output $\boldsymbol{h}_t$ at the last time step $t$ can be served as the user's current consumption motivation representation $\boldsymbol{h}_{\mathrm{SBL}}$.
\textcolor{black}{In PBL, BINN adopts a bidirectional CLSTM (Bi-CLSTM) which considers both forward and backward input sequences to obtain the long-term preference representation $\boldsymbol{h}_{\mathrm{PBL}}$. By concatenating $\boldsymbol{h}_{\mathrm{SBL}}$ with $\boldsymbol{h}_{\mathrm{PBL}}$, the obtained representation is utilized to make predictions and generate recommended items.}
\textcolor{black}{BINN proposes a novel gating structure, consisting of Bi-CLSTM and CLSTM, which enables the memorization of multi-behavior information in sequences. In contrast to RLBL and RIB in how to introduce multi-behavior information, BINN modifies the internal structure of the LSTM by feeding behavior embedding matrix into the Bi-CLSTM and CLSTM to make it suitable for multi-behavior sequences. However, the limitations of BINN are similar to those of RIB, as both represent multiple types of user behavior directly in embedding matrix, which might make it challenging to capture real user behavioral information.}

\textbf{IARS.}
Intention-aware recommender system (IARS)~\cite{TOIS2021-IARS} is also a work that incorporates the item category \textcolor{black}{to perform the next-item recommendation task}. IARS consists of four blocks in total, which are an RNN-based encoder for perceiving user intent, and three decoders, i.e., a judgment or prediction task based on user intent, so as to learn the complex and co-existing intent of the user. Specifically, the encoder takes a sequence of (item, behavior, category) tuples as input, adopts a local modeling perspective, processes the behavior types through
\textcolor{black}{multiple multi-behavior GRU units (MGRUs) to capture multiple intentions of the user. Note that we only discuss Soft-MGRU, one type of MGRU, for its lower spatial complexity and better performance by sharing the same set of parameters between different behaviors. After an embedding layer, the item embedding, category embedding and behavior embedding are fed into Soft-MGRU to obtain the hidden state $\boldsymbol{h}_{t}$ at the time step $t$.}

Soft-MGRU encodes the dependencies of items in multi-behavior sequences and obtains hidden states that characterize user intentions. 
It takes into account item categories and utilizes an attention network to capture the user's purchase intention for candidate items. The introduction of multi-behavior information is also achieved through the GRU's input which concatenates the embeddings of behavior, item, and category. However, the behavior embedding only participates in the computation of the reset gate and update gate, which may not be sufficient to represent the complete behavior information of the user.

\textbf{MBN.}
Multi-behavior network (MBN)~\cite{TKDD2022-MBN} models multi-behavior sequences towards the next-basket recommendation problem.
The MBN architecture is composed of three modules, i.e., basket encoder, meta multi-behavior sequence encoder and recurring-item-aware predictor. 
\textcolor{black}{Specifically, the basket encoder converts the item representation $\boldsymbol{e}_{v}$ to the basket representation of the items $\boldsymbol{E}_{u, t}^{f}$ by a max pooling method.}
In the meta multi-behavior sequence encoder, multiple behavior-specific subsequences of items are taken as input and go through Behavior-RNN layers to learn behavior-specific information, which is local in the perspective of modeling. In addition to the Behavior-RNN layers, this work also proposes a Meta-RNN layer to learn the collective knowledge of \textcolor{black}{multi-behavior} sequences. Then a gathering-scattering scheme is utilized to correlate the Meta-RNN layer and the Behavior-RNN layer. The representations learned by the Behavior-RNN layers are gathered to the Meta-RNN layer to learn the collective knowledge of \textcolor{black}{multi-behavior} sequences, and then the representations learned by the Meta-RNN layer are scattered to the individual Behavior-RNN layer to calibrate behavioral modeling.
In the recurring-item-aware predictor, a mixed probabilistic function in the generate mode and the repeat mode is proposed to predict the probability of each item in the next basket, which can simulate the distribution of items with biased repetition.

MBN introduces a method of gathering and then scattering to fuse and assign the learned multi-behavior information to different Behavior-RNNs layers at the Meta RNN layer, which is a more explicit way to model intra-behavioral and inter-behavioral sequence information. \textcolor{black}{In addition, any type of user's behavior can be treated as the target behavior.} Nonetheless, as the number of behavior types increases, the number of behavioral RNN layers and associated parameters also increases, resulting in heightened computational complexity. Moreover, the division of the item basket in MBN is based on the time span, which may not align with the real-world scenario of purchasing a basket of items at the same time.

\textcolor{black}{In addition to the above works, several efforts employ RNN-based learning architecture to model the sequentiality and heterogeneity of user behaviors. CBS~\cite{IJCAI2018-CBS} models longer sequences rather than short-term dependencies for the next-basket recommendation problem with the use of a LSTM with or without shared parameters for each of the two behaviors (or the representation obtained from the embedding layer directly for the target behavior sequence). DIPN~\cite{KDD2019-DIPN} employs a GRU and a hierarchical attention mechanism to effectively capture heterogeneous user behaviors and utilizes a multi-task module to capture short-term and long-term purchase preferences. HUP utilizes the attention mechanism, and designs LSTMs with the addition of behavior gate and time gate at the micro-, item-, and category-levels to capture different granularities of information from session-based recommendation. In terms of federated recommendation, DeepRec~\cite{WWW2021-DeepRec} applies GRU on the historical interaction data of all users on the cloud, and is then pushed to users' devices, which makes it possible to fine-tune it for the individuals to obtain a personal recommendation model for each of them.}

\textcolor{black}{In summary, the} RNN-based learning architecture is suitable for sequence problems and can store short-term memories, but suffers from gradient disappearance and gradient explosion problems. \textcolor{black}{In addition, RNN is inefficient and has difficulty in predicting information about future sequences since the output of the current moment depends on the computation and the output of the previous moment.} At present, the industry has \textcolor{black}{rarely} leveraged RNN-based learning architecture for recommendation.

\subsection{GNN-based Learning Architecture}
Graph neural network (GNN)~\cite{TNN2008-GNN,wu2020gnn}, utilized to extract features, is a widespread technique in recent years, and \textcolor{black}{there have been many excellent graph neural network models, including GCN~\cite{ICLR2016-GCN}, GraphSAGE~\cite{NeurIPS2017-GraphSAGE}, GAT~\cite{ICLR2018-GAT} and so on}. It can fully exploit the higher-order neighbor information of nodes and performs well on recommender systems.

\subsubsection{Basic Paradigm}
\begin{figure}[!htb]
	
	\begin{center}
		
		\begin{tabular}{cc}
			
			\includegraphics[width=1.5in]{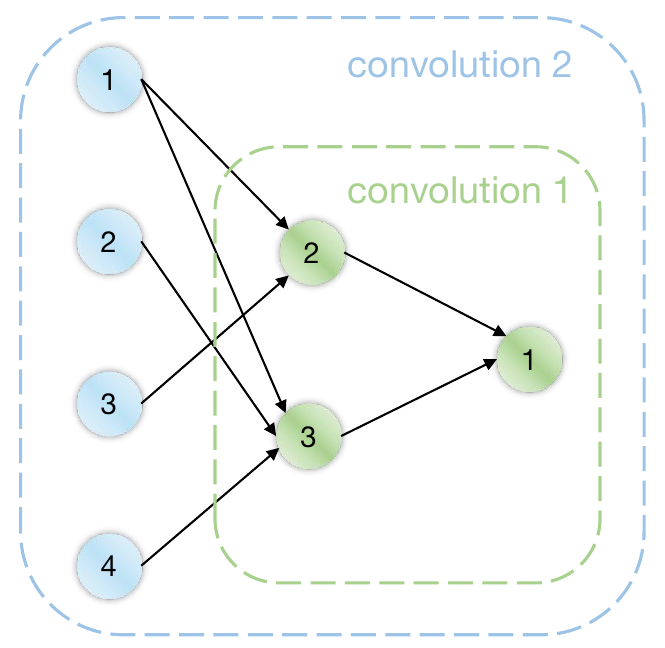}
			
		\end{tabular}
		
	\end{center}
	
	\caption{An example of graph convolution.}
	
	\label{fig:Graph}
\end{figure}

In general, graph neural network models use graph convolution to allow nodes to obtain information about their neighbors. \textcolor{black}{To make the procedure more specific, an example} is shown in Figure~\ref{fig:Graph}, which depicts four nodes, labeled as node 1, node 2, node 3, and node 4. Node 1's first-order neighbors are node 2 and node 3. During the first-order graph convolution, the embeddings of node 2 and node 3 are aggregated into the embedding of node 1. In the second-order graph convolution, node 3 is a neighbor of node 4. Since node 3 has already obtained information about node 4 in the first-order graph convolution, node 1 is able to obtain information about its second-order neighbor, node 4, during the second-order graph convolution. This allows the graph convolution network to effectively utilize information from higher-order neighbors of nodes.
\begin{table*}[htbp]

\caption{Data \& modeling perspectives and features used in works based on GNN learning architecture.} \label{tbl:GNN}
\begin{center}
\small
\begin{tabular}{  p{1.8cm} | p{3cm} | p{2.5cm} |  p{9cm}} \hline\hline

Works & Data Perspective & Model Perspective & Features\\
    \hline\hline
MGNN-SPred~\cite{WWW2020-MGNN-Spred}
    & Some behavior-specific subsequences of items
    & Global
    & Modeling behavior from behavior transition relations, containing \textcolor{black}{homogeneous behavior transitions intra each kind of behavior-specific subsequences}.\\
    \hline
DMBGN~\cite{KDD2021-DMBGN}
    & Some behavior-specific subsequences of items
    & Global
    & Focus on the task of voucher redemption rate prediction and \textcolor{black}{model the relationship between} multiple behaviors and vouchers effectively.\\
    \hline
GPG4HSR~\cite{RecSys2022-GPG4HSR}
    & A sequence of (item, behavior) pairs
    & Local + Global
    & Learn various behavior transition relations from the global graph and the personalized graph, respectively.\\
    \hline
BGNN~\cite{CS2022-BGNN}
    & Some behavior-specific subsequences of items
    & Global
    & Construct directed graphs for different behavior transition (homogeneous and heterogeneous) information.\\
    \hline    
BA-GNN~\cite{CS2022-BA-GNN}
    & Some behavior-specific subsequence of items
    & Global
    & Construct directed graphs for different behavior-specific sequences respectively.\\
\hline\hline
\end{tabular}
\end{center}
\end{table*}
\subsubsection{Methods in MBSR}
In MBSR, there are lots of works achieving great recommendation performance based on GNN, such as MGNN-SPred~\cite{WWW2020-MGNN-Spred}, DMBGN~\cite{KDD2021-DMBGN}, GPG4HSR~\cite{RecSys2022-GPG4HSR}, BGNN~\cite{CS2022-BGNN} and BA-GNN~\cite{CS2022-BA-GNN}. We describe \textcolor{black}{some of them} in detail below, and summarize the data perspective, the modeling perspective and the characteristics of these works in Table~\ref{tbl:GNN}.

\textbf{MGNN-SPred.}
Multi-relational graph neural network model for session-based target behavior prediction (MGNN-SPred)~\cite{WWW2020-MGNN-Spred} also utilizes \textcolor{black}{GNN} to model \textcolor{black}{multi-behavior} sequences \textcolor{black}{in session-based recommendation scenarios} from a global modeling perspective. 

Firstly, MGNN-SPred treats the \textcolor{black}{multi-behavior} sequences $\mathcal{S}_{\mathrm{p}}$ and $\mathcal{S}_{\mathrm{e}}$ as some sequences of behavior-specific items, and constructs a global graph from all the training sequences, where the nodes represent items, and the edges have two attributes, namely, \textcolor{black}{purchase} edges and examination edges. For example, the purchase edge of item $a$ and item $b$ means a user purchases item $a$ and then purchases item $b$.
For each node $v$, we can obtain four types of neighbor node subsets, i.e.,  $\mathcal{N}{\textcolor{black}{_{\mathrm{p+}}(v)}}$, $\mathcal{N}{\textcolor{black}{_{\mathrm{e+}}(v)}}$, $\mathcal{N}{\textcolor{black}{_{\mathrm{p-}}(v)}}$ and $\mathcal{N}{\textcolor{black}{_{\mathrm{e+}}(v)}}$. For example, $\mathcal{N}{\textcolor{black}{_{\mathrm{e+}}(v)}}$ and $\mathcal{N}{\textcolor{black}{_{\mathrm{e-}}(v)}}$ \textcolor{black}{denote the incoming edges and outgoing edges of the node which is treated as an examined item, respectively. We give the concrete forms of the neighbor node subsets of the node $v$ as follows:}
\begin{align}
    &\mathcal{N}{_{\mathrm{p+}}(v)}=\left\{v^{\prime}|\left(v^{\prime}\to v,\mbox{purchase}\right)\in \mathcal{E}\right\}\\
    &\mathcal{N}{_{\mathrm{e+}}(v)}=\left\{v^{\prime}|\left(v^{\prime}\to v,\mbox{examination}\right)\in \mathcal{E}\right\}\\
    &\mathcal{N}{_{\mathrm{p-}}(v)}=\left\{v^{\prime}|\left(v\to v^{\prime},\mbox{purchase}\right)\in \mathcal{E}\right\}\\
    &\mathcal{N}{_{\mathrm{e-}}(v)}=\left\{v^{\prime}|\left(v\to v^{\prime},\mbox{examination}\right)\in \mathcal{E}\right\}
\end{align}
where $\mathcal{E}$ denotes the edge set.
Secondly, for a target item $v$, the k-level aggregated representations of four different neighbors (taking $\mathcal{N}{_{\mathrm{p+}}(v)}$ as an example) and the node representation ${\boldsymbol{h}}_{v}^{k}$ obtained from the final iteration can be calculated as follows:
\begin{align}
&\boldsymbol{h}_{\mathrm{p+}, v}^{k}=\frac{\sum_{v^{\prime} \in \mathcal{N}_{\mathrm{p+}}(v)} \boldsymbol{h}_{v^{\prime}}^{k-1}}{\left|\mathcal{N}_{\mathrm{p+}}(v)\right|} \\
&{\boldsymbol{h}}_{v}^{k}={\boldsymbol{h}}_{v}^{k-1}+\boldsymbol{h}_{\mathrm{p+}, v}^{k}+\boldsymbol{h}_{\mathrm{e+}, v}^{k}+\boldsymbol{h}_{\mathrm{p-}, v}^{k}+\boldsymbol{h}_{\mathrm{e-}, v}^{k}
\end{align}
where ${\boldsymbol{h}}_{v}^{k}   \in  \mathbb{R}^{d \times 1}$ and ${\boldsymbol{h}}_{v}^{K}   \in  \mathbb{R}^{d \times 1}$ denote the $k$-th step and  the last step of the item $v$ representation in GNN, respectively.
\textcolor{black}{And ${\boldsymbol{h}}_{v}^{K}$ is used as the corresponding item potential representation.}
Thirdly, it treats the multi-behavior sequences as some sequences of behavior-specific items and \textcolor{black}{obtains the user's examination preference and purchase preference by aggregating all item potential representations of the examination sequence  and the purchase sequence, respectively. The final preference representation is obtained after feeding the above two preferences to the fully connected layer and the gated network.}

MGNN-SPred is a simple but effective method for GNN to be directly applied to MBSR. By constructing a graph, the sequential occurrence relationships of different behaviors are reflected in the graph, allowing aggregation between different behaviors and enhancing their information capability. Additionally, the MGNN-SPred approach, which first models distinct behavioral sequences individually before passing through the gated neural network, effectively captures both intra-behavioral and inter-behavioral information, ensuring a well-balanced information representation.
\textcolor{black}{Due to these advantages of MGNN-SPred, some of the subsequent works for MBSR are based on MGNN-SPred with some improvements~\cite{CS2022-BGNN,CS2022-BA-GNN,MLBDBI2021-MBGNN}. For example, the improvement of BA-GNN~\cite{CS2022-BA-GNN} over MGNN-SPred is that BA-GNN constructs separate graphs for different behavior sequences, and utilizes a gated graph neural network (GG-NNs)~\cite{ICLR2016-GGSNN} and a sparse self-attention mechanism to address the noise effect in the examination sequence, thus better capturing the information in multi-behavior sequences.}

\textbf{DMBGN.}
Deep multi-behavior graph networks (DMBGN)~\cite{KDD2021-DMBGN} focuses on the task of voucher redemption rate prediction \textcolor{black}{in the session-based recommendation scenario}.
It utilizes GNN to model \textcolor{black}{users' long-term voucher redemption preferences} from a global perspective.

Firstly, it treats the multi-behavior sequences $\mathcal{S}_{\mathrm{atc}}$ and $\mathcal{S}_{\mathrm{ord}}$ as some sequences of behavior-specific items, and divides them into four parts, i.e., $\mathcal{S}_{\mathrm{atc}}^{\mathrm{+}}$, $\mathcal{S}_{\mathrm{atc}}^{\mathrm{-}}$, $\mathcal{S}_{\mathrm{ord}}^{\mathrm{+}}$ and $\mathcal{S}_{\mathrm{ord}}^{\mathrm{-}}$. For example, the sequence $\mathcal{S}_{\mathrm{atc}}^{\mathrm{+}}$  means that the behaviors of add-to-cart happen before the behavior on the voucher, while the sequence $\mathcal{S}_{\mathrm{atc}}^{\mathrm{-}}$  means that the behaviors of add-to-cart happen after the behavior on the voucher. $\mathcal{S}_{\mathrm{atc}}^{\mathrm{+}}$, $\mathcal{S}_{\mathrm{atc}}^{\mathrm{-}}$, $\mathcal{S}_{\mathrm{ord}}^{\mathrm{+}}$ and $\mathcal{S}_{\mathrm{ord}}^{\mathrm{-}}$ are connected to the central voucher node by the closest items from the temporal perspective. We obtain four sub-graphs in the end, i.e., $\mathrm{atc}_{\mathrm{+}}$, $\mathrm{atc}_{\mathrm{-}}$, $\mathrm{ord}_{\mathrm{+}}$ and $\mathrm{ord}_{\mathrm{-}}$.
Secondly, the four sub-graphs constructed above are fed into \textcolor{black}{GNN} with the Weisfeiler-Leman algorithm~\cite{morris2019weisfeiler}, separately.
The representations of the four sequences $\mathcal{S}_{\mathrm{atc}}^{\mathrm{+}}$, $\mathcal{S}_{\mathrm{atc}}^{\mathrm{-}}$, $\mathcal{S}_{\mathrm{ord}}^{\mathrm{+}}$ and $\mathcal{S}_{\mathrm{ord}}^{\mathrm{-}}$ are concatenated and \textcolor{black}{sent to} a multilayer perceptron (MLP) function.
\textcolor{black}{Then the final UVG embedding is generated by concatenating the output of the MLP function and the embedding of the central voucher node, and the score for the historical UVG is obtained by the dot-product between the two representations. Finally,} the representation of the embedding calculated from the target UVG component \textcolor{black}{is enhanced via an attention network.}. 

GNN can \textcolor{black}{model the relationship between} multiple behaviors and vouchers effectively. In building the graph, it is also reasonable that the coupon node is only connected to nodes of other behaviors that are temporally close, which improves the relationship between temporally close nodes and the coupon.  \textcolor{black}{Furthermore, DMBGN incorporates all historical sequences into the GNN network, thus proficiently capturing users' long-term preferences. The output of these past sequences is subjected to attention, together with the output of the current sequence, effectively enhancing the information representation of the current sequence.}

\begin{figure}[!htb]
	
	\begin{center}
		
		\begin{tabular}{cc}
			
			\includegraphics[width=3.2in]{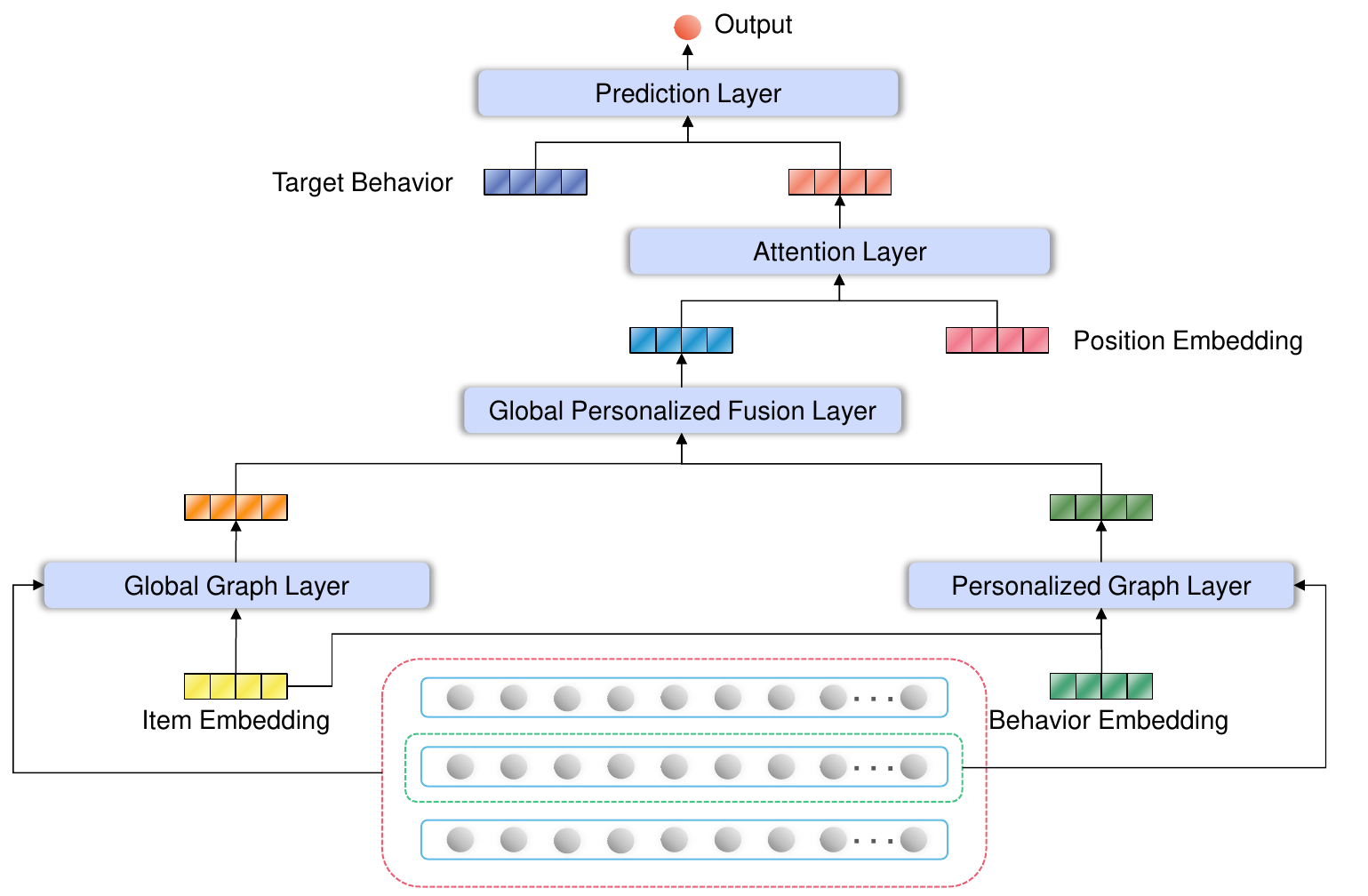}
			
		\end{tabular}
		
	\end{center}
	
	\caption{Illustration of global and personalized graphs for heterogeneous sequential recommendation (GPG4HSR).}
	
	\label{fig3:GPG4HSR}
\end{figure}

\textbf{GPG4HSR.}
Global and personalized graphs for heterogeneous sequential recommendation (GPG4HSR)~\cite{RecSys2022-GPG4HSR} simultaneously considers the \textcolor{black}{transition} relationships between different behaviors and local contextual information, thereby improving the \textcolor{black}{next-item} recommendation performance, which \textcolor{black}{we} illustrate in Figure~\ref{fig3:GPG4HSR}. 
GPG4SHR focuses on two types of behaviors, i.e., examinations and purchases, and takes a sequence of (item, behavior) pairs as input to model all sequences and each sequence from a global and local perspective, respectively. 
\textcolor{black}{Specifically, GPG4HSR first feeds the input into an embedding layer to obtain the item embedding $\boldsymbol{v}_{i^t}$, the behavior embedding $F_{f^t}$ and the position embedding $\boldsymbol{p}_{t}$ of the item $i^t$ interacted by the behavior type $f^t$ at the time step $t$. Then the embeddings}
are introduced to a global graph layer and a personalization layer to capture the transition patterns between behaviors and users' intent considering adjacent contextual information, respectively. In the global graph layer,
\textcolor{black}{the input is the global node $v_{i^t}$ of each item of a sequence (abbreviated as $v$) and all the edges linked to it in the global graph, where there are six edge types to distinguish specific behavioral transitions,}
(i.e., $e 2 e$, $p 2 p$, $e 2 p+$, $e 2 p-$, $p 2 e+$ and $p 2 e-$) that considers the transition directions (inward or outward) between different behaviors on top of TransRec++. The
corresponding neighbor node subsets are $\mathcal{N}{_{\mathrm{e2e}}(v)}$, $\mathcal{N}_{\mathrm{p2p}}{(v)}$, $\mathcal{N}_{\mathrm{e2p+}}{(v)}$, $\mathcal{N}_{\mathrm{e2p-}}{(v)}$, $\mathcal{N}_{\mathrm{p2e+}}{(v)}$ and $\mathcal{N}_{\mathrm{p2e-}}{(v)}$. 
The final generated neighbor group and \textcolor{black}{behavior transition-specific representation} are as follows (take $e2p+$ as an example):
\begin{align}
    &\mathcal{N}{_{\mathrm{e2p+}}(v)}=\left\{\left(v^{\prime},\mathrm{freq}\right)|\left(v^{\prime}\to v,\mathrm{freq, \mathrm{e2p+}}\right)\in \mathcal{E}_{\mathrm{g}}\right\}\\
    &\boldsymbol{h}_{v}^{\mathrm{e2p+}}=\frac{\sum_{(v^{\prime},\mathrm{freq}) \in \mathcal{N}_{\mathrm{e2p+}}(v)} \mathrm{freq} \times \boldsymbol{v}_{v^{\prime}}}{\sum_{(v^{\prime}, \mathrm{freq}) \in \mathcal{N}_{\mathrm{e2p+}}(v)} \mathrm{freq}}
\end{align}
\textcolor{black}{where $\boldsymbol{v}_{v^{\prime}}$ is a concise representation of the node $v^{\prime}$ linked to the node $v$, $\mbox{freq} \in \mathbb{R}$ represents the frequency of the corresponding edge. Then the global graph representation $\boldsymbol{h}_{v}^{g}$ of the node $v$ can be represented as the sum of the node representation with the weighted representation of all behavior transition representations.}

\textcolor{black}{In the personalization graph layer, the input contains the item embedding and the behavior embedding of the node $v$, i.e., $\left(\boldsymbol{v}_{v}+F_{v}\right)$, as well as the inward and outward attributes of node-connected edges, which can learn the importance of different behaviors and the adjacent context information, thereby capturing users’ intent.}
The final graph representation $\boldsymbol{h}_{v}$ of the node $v$ obtained by fusing the global graph representation $\boldsymbol{h}_{v}^{g}$ and the personalized graph representation $\boldsymbol{h}_{v}^{u}$:
\begin{align}
    \boldsymbol{h}_{v}^{u}=\boldsymbol{v}_{v}+\textcolor{black}{F_{v}}+\boldsymbol{h}_{v}^{+}+\boldsymbol{h}_{v}^{-}\\
    \boldsymbol{h}_{v}=\gamma_{u} \boldsymbol{h}_{v}^{g}+\left(1-\gamma_{u}\right) \boldsymbol{h}_{v}^{u}
\end{align}
where $\gamma_{u}=\sigma\left(\boldsymbol{W}_{g p}\left[\boldsymbol{h}_{v}^{g} ; \boldsymbol{h}_{v}^{u}\right]\right)$. The final graph representation of the sequence $\boldsymbol{h}_{u}$ can be obtained by concatenating the graph representation of the corresponding nodes of the sequence,
\textcolor{black}{and then passed into a dropout layer and stacked self-attention blocks same as SASRec~\cite{ICDM2018-SASRec} together with the corresponding position embedding. The obtained representation is concatenated with the target behavior vector and fed into a softmax function to obtain the user's predicted preference value for the items.}

GPG4HSR constructs both a global graph and a personalized graph, where the global graph is used to capture the relationships among heterogeneous behaviors, and the personalized graph is used to enhance the contextual representation of a single user's multi-behavior sequence for a better comprehension of the user’s preferences. 
\textcolor{black}{In addition, the graph construction with the time complexity $O(|\mathcal{R}|)$, where $\mathcal{R}$ denotes the set of the user-item interaction, making it more efficient.}
Nevertheless, as the number of behavior types increases, the number of behavior transfer relationship types also increases, which increases the complexity of graph construction. Moreover, the multi-order behavior transition relationships, which typically optimize performance, can pose challenges in modeling.

\textbf{BGNN.}
Behavior-aware graph neural network (BGNN)~\cite{CS2022-BGNN} distinguishes between two different behavioral sequences by utilizing a dual-channel learning strategy \textcolor{black}{for the session-based recommendation.} BGNN takes an examination sequence and a purchase sequence as input, and models the heterogeneous behavior transitions to obtain the semantic connections between diverse behaviors by two global graphs, i.e., homogeneous behavior transition graph (HoBTG) and heterogeneous behavior transition graph (HeBTG), so as to improve the recommendation performance.

Specifically, BGNN sends \textcolor{black}{the examination sequence and the purchase sequence into the auxiliary channel and the target channel, respectively.} In the target channel, the item representation is learned in the purchase sequence through HoBTG, \textcolor{black}{which is basically equivalent to the modeling of the behavior transition relationship of MGNN-SPred.}
\textcolor{black}{The auxiliary channel consists of three modules to learn the item presentation of the examination sequence. The first module directly uses homogeneous behavior transition in the target channel to obtain potential representation; the second module adaptively adjusts the contributions of different neighbors of nodes through an attention mechanism to learn the purchase-oriented item representation; and the third module is for representation aggregation, which is the representation of items obtained by balancing the above two modules by gathering.}
After obtaining the item presentation matrices of the examination sequence and the purchase sequence, the matrices are sent to an attention network separately and then fused together. Finally, the user's preference value for the item is obtained through a prediction layer.

\textcolor{black}{BGNN constructs graphs that explicitly capture two behavior transition patterns of homogeneous and heterogeneous ones, and utilizes these graphs in the auxiliary behavior to capture the contribution of the auxiliary behavior to the target behavior, thereby improving the user's next preferred item prediction under the target behavior, \textcolor{black}{though its training time is about 1.4 times of that of MGNN-SPred}. However, BGNN encounters difficulties in the setting of more behavior types, that is, the transition relationships between behaviors become more complex with an increase in the number of behavior types, and additional graphs may also need to be constructed, thus increasing the complexity of the algorithm.}

\textcolor{black}{In summary, by} constructing the user-item graph, \textcolor{black}{GNN} can be easily applied to the recommendation system methods. For each graph node, the aggregation of neighbors' information allows each behavior to obtain information about other behaviors that occurred close in time. The recommendation performance is obviously enhanced by the information enhancement of the neighbor nodes. \textcolor{black}{In comparison to RNN, GNN has the ability to model more complex relationships within multi-behavior sequences, and possesses stronger capabilities to handle data sparsity.}

\subsection{Transformer-based Learning Architecture}
The Transformer model~\cite{NIPS2017Transformer}, \textcolor{black}{a deep learning architecture that utilizes a self-attention network to reduce the computational complexity and thus enhance the training speed, has gained wide recognition in recent years for its superior performance in sequence-to-sequence modeling. This model has been utilized in a wide range of research areas, including} natural language processing~\cite{devlin2018bert,liu2019roberta}, computer vision~\cite{dosovitskiy2020image,carion2020end}, and recommender systems~\cite{CIKM2020-DMT,IJCAI2020-DFN,TIST2022-FLAG}.
\subsubsection{Basic Paradigm}
\label{Transformer Basic Paradigm}
\begin{figure}[!htb]
	
	\begin{center}
		
		\begin{tabular}{cc}
			
			\includegraphics[width=3.2in]{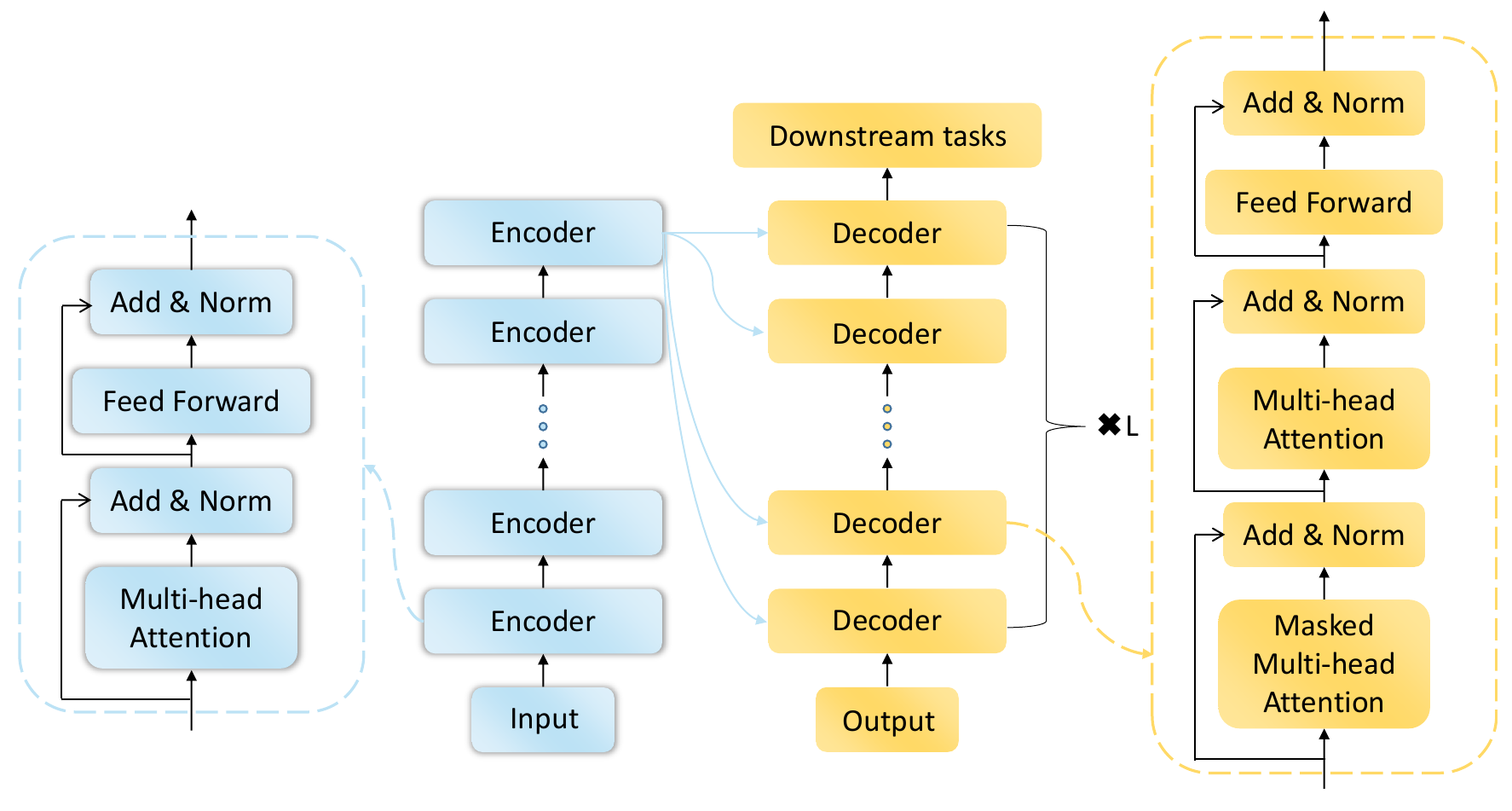}
			
		\end{tabular}
		
	\end{center}
	
	\caption{Illustration of Transformer.}
	
	\label{fig:Transformer}
\end{figure}
The basic architecture of \textcolor{black}{the Transformer model is depicted} in Figure~\ref{fig:Transformer}. It consists of two modules: the encoders and the decoders. In this discussion, we will focus on the encoders. \textcolor{black}{The most crucial component within the encoder is the multi-head self-attention component.} This component comprises several self-attention subcomponents, which are widely used in recommendation models. We will specifically examine the self-attention component by considering the representation of an examination sequence $\mathcal{S}_{\mathrm{e}}$. The calculation of the self-attention component is as follows:
\begin{align}
	&\boldsymbol{Q}_i= \mbox{Em}(\mathcal{S}_{\mathrm{e}}        )\boldsymbol{W}_i^{Q}, \quad \boldsymbol{K}_i= \mbox{Em}(\mathcal{S}_{\mathrm{e}})\boldsymbol{W}_i^{K}, \quad \boldsymbol{V}_{i}=\mbox{Em}(\mathcal{S}_{\mathrm{e}} )\boldsymbol{W}_i^{V}\\
	&\boldsymbol{head}_{i}=\mbox{softmax}\left(\frac{\boldsymbol{Q}_i^{\top} \boldsymbol{K}_i}{\sqrt{d_t}}\right) \boldsymbol{V}_{i}\\
	&\boldsymbol{F}_{\mathrm{e}}=\mbox{concatenate}\left(\boldsymbol{head}_{1}, \cdots, \boldsymbol{head}_{h}\right)  \boldsymbol{W}^{O}
\end{align}
where $ \mbox{Em}(\mathcal{S}_{\mathrm{e}})  \in \mathbb{R}^{n_\mathrm{e} \times d}, \boldsymbol{W}_i^{Q} \in \mathbb{R}^{d\times d_{t}}$, $\boldsymbol{W}_i^{K} \in \mathbb{R}^{d\times d_{t}}$, $\boldsymbol{W}_i^{V} \in \mathbb{R}^{d\times d_{t}}$ and $\boldsymbol{W}^{O} \in \mathbb{R}^{{hd}_t\times d}$ are  the projection matrices. $n_\mathrm{e}$ is the length of the sequence $\mathcal{S}_{\mathrm{e}}$, $d_{t}$ is the dimension of  $ \boldsymbol{K}_i$, and \textcolor{black}{$\boldsymbol{F}_{\mathrm{e}}$ is the output of the self-attention component.}

\subsubsection{Methods in MBSR}
In MBSR, there are some works obtaining great recommendation performance based on Transformer, including DMT~\cite{CIKM2020-DMT}, DFN~\cite{IJCAI2020-DFN}, FeedRec~\cite{WWW2022-FeedRec}, NextIP~\cite{CIKM2022-NextIP}, MB-STR~\cite{SIGIR2022-MB-STR}, DUMN~\cite{RecSys2021-DUMN} and FLAG~\cite{TIST2022-FLAG}. We describe \textcolor{black}{some of them} in detail below, and summarize the data perspective, the modeling perspective and characteristics of these works in Table~\ref{tbl:Transformer}.

\begin{figure*}[!htb]
	
	\begin{center}
		
		\begin{tabular}{cc}
			
			\includegraphics[width=5.2in]{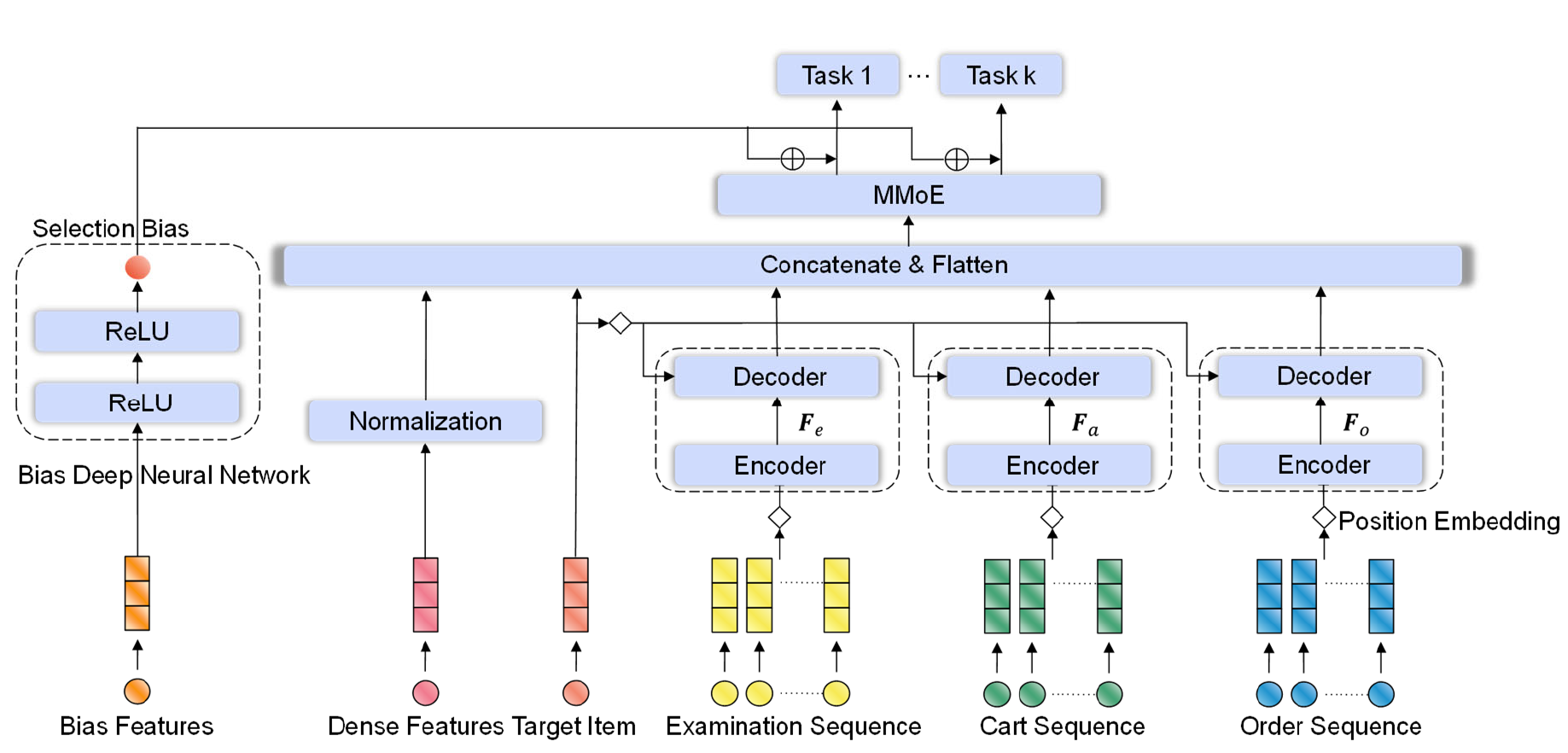}
			
		\end{tabular}
		
	\end{center}
	
	\caption{Illustration of deep multifaceted Transformers (DMT).}
	
	\label{fig:DMT}
\end{figure*}

\begin{table*}[htbp]
\caption{Data \& modeling perspectives and features used in works based on Transformer learning architecture.} \label{tbl:Transformer}
\begin{center}
\small
\begin{tabular}{ p{1.5cm} | p{4.3cm} | p{2.5cm} |  p{8cm}} \hline\hline

Works & Data Perspective & Model Perspective & Features\\
    \hline\hline
DMT~\cite{CIKM2020-DMT}
    & Some behavior-specific subsequences of items
    & \textcolor{black}{Local + Global}
    &  \textcolor{black}{Use target item as query; Consider implicit feedback bias by a bias deep neural network.}\\
    \hline
DFN~\cite{IJCAI2020-DFN}
    & Some behavior-specific subsequences of items
    & \textcolor{black}{Local + Global}
    & \textcolor{black}{Use target item as query; Consider implicit negative feedback noise by an attention network .}\\
    \hline
DUMN~\cite{RecSys2021-DUMN}
    & Some behavior-specific subsequences of items
    & Local
    & \textcolor{black}{Consider implicit feedback \textcolor{black}{noise}; Use memory network to obtain the long-term user preference.}\\
    \hline
FeedRec~\cite{WWW2022-FeedRec}
    & Some behavior-specific subsequences of items \textcolor{black}{and a sequence of (item, behavior) pairs}
    & Local + Global 
    & \textcolor{black}{Consider implicit feedback \textcolor{black}{noise} by an attention network; Consider multiple patterns of the multi-behavior sequences.}\\
    \hline
NextIP~\cite{CIKM2022-NextIP}
    & \textcolor{black}{Some behavior-specific subsequences of items and a sequence of (item, behavior) pairs}
    & \textcolor{black}{Local + Global}
    & \textcolor{black}{Treat the problem as the item prediction task and the purchase prediction task; Consider multiple patterns of the multi-behavior sequences.} \\
    \hline
MB-STR~\cite{SIGIR2022-MB-STR}
    & A sequence of (item, behavior) pairs
    & \textcolor{black}{Local}
    & A novel positional encoding function to model \textcolor{black}{multi-behavior} sequence relationships.\\
    \hline
FLAG~\cite{TIST2022-FLAG}
    & A behavior-agnostic sequence of items and \textcolor{black}{a sequence of behaviors}
    & Local + Global
    & Model user’s local preference, local intention and global preference simultaneously.\\
\hline\hline 

\end{tabular}

\end{center}
\end{table*}
\textbf{DMT.}
Deep Multifaceted Transformers (DMT)\cite{CIKM2020-DMT} utilizes a multi-gate mixture-of-experts (MMoE) \textcolor{black}{approach, a multi-task learning technique, to enhance the performance of both click-through rate (CTR) and click value rate (CVR) predictions.} Furthermore, it employs the Transformer model to analyze multi-behavior sequences from a local modeling perspective. The architecture of DMT is depicted in Figure\ref{fig:DMT}.

Firstly, it treats the multi-behavior sequences as some sequences of behavior-specific items and inputs those into the encoder of Transformer. 
Take the examination sequence as an example, and the formula is as follows:
\begin{align}
	&\boldsymbol{F}_{\mathrm{e}} = \mbox{Encoder}\left(\mbox{PE}\left(\mbox{Em}( \mathcal{S}_\mathrm{e})\right)\right)\\
	&\boldsymbol{E}_{\mathrm{e}} = \mbox{Decoder}(\boldsymbol{F}_{\mathrm{e}},\mbox{PE}(V_{i}^{\mathrm{target}}))
\end{align}
where $\mbox{PE}(\cdot)$ is the positional encoding function, and it explicitly represents the sinusoidal positional embedding or the learned positional embedding \cite{ICDM2018-SASRec, NIPS2017Transformer} in DMT. $V_{i}^{\mathrm{target}}$ is the embedding of the item to be predicted and \textcolor{black}{one of the inputs for the decoder of Transformer.}
Secondly,  $\boldsymbol{E}_{\mathrm{e}}, \boldsymbol{E}_{\mathrm{a}}$ and $\boldsymbol{E}_{\mathrm{o}}$ are concatenated and flattened with the normalized dense features gathered from the recommender system and the target item embedding $V_{i}^{\mathrm{target}}$.
Thirdly, the multi-task training model MMoE is used to improve the performance of both CTR and CVR prediction. In particular, DMT considers bias in implicit feedback, such as position and neighboring bias, \textcolor{black}{and utilizes a deep neural network with the ReLU function.} 


DMT uses a Transformer with unshared parameters to capture the relationships within each behavior and subsequently feeds the different behavioral features into the MMoE module, lacking the explicit modeling of the relationships between the different behaviors.  A bias deep neural network is proposed for modeling implicit feedback bias, which is a good modeling solution.

\textbf{DFN.}
Deep feedback network (DFN)~\cite{IJCAI2020-DFN}, \textcolor{black}{another work for CTR prediction in ads}, models multi-behavior sequences utilizing Transformer from a local modeling perspective and three modules commonly used in industry, i.e., a wide component, an FM component, and a deep component.

\textcolor{black}{We can draw a comparison between DFN and DMT~\cite{CIKM2020-DMT}. Firstly, like DMT, DFN employs a Transformer architecture with unshared parameters to capture the relationships within each behavior. It treats} multi-behavior sequences as a series of behavior-specific items and inputs them into a multi-head self-attention mechanism. \textcolor{black}{Secondly, DFN also takes into account the implicit feedback \textcolor{black}{noise}. Unlike DMT, DFN leverages the attention mechanism to explore  the relationship between different behaviors, which can be advantageous.}
Noting that the implicit negative feedback, i.e., the unexamination sequence $\mathcal{S}_{\mathrm{n}}$, is abundant in real life but contains noise. As such, DFN uses implicit positive feedback $\boldsymbol{f}_{\mathrm{e}}$ and explicit negative feedback $\boldsymbol{f}_{\mathrm{d}}$ to denoise the implicit negative feedback by an attention network.  The formula is as follows:
\begin{align}
	&\boldsymbol{f}_{\mathrm{ne}} = \mbox{attention}(\mbox{Em}(\mathcal{S}_{\mathrm{n}}),\boldsymbol{f}_{\mathrm{e}})\\
	&\boldsymbol{f}_{\mathrm{nd}} = \mbox{attention}(\mbox{Em}(\mathcal{S}_{\mathrm{n}}),\boldsymbol{f}_{\mathrm{d}})
\end{align} 
where $\boldsymbol{f}_{\mathrm{e}}  \in \mathbb{R}^{1 \times d}$ and $\boldsymbol{f}_{\mathrm{d}}   \in \mathbb{R}^{1 \times d}$ are the keys, and  $\boldsymbol{f}_{\mathrm{ne}}   \in \mathbb{R}^{1 \times d}$ and  $\boldsymbol{f}_{\mathrm{nd}}   \in \mathbb{R}^{1 \times d}$ are the outputs of the two attention networks, respectively.
Finally,  $\boldsymbol{f}_{\mathrm{e}}$, $\boldsymbol{f}_{\mathrm{d}}$, $\boldsymbol{f}_{\mathrm{n}}$, $\boldsymbol{f}_{\mathrm{nc}}$ and  $\boldsymbol{f}_{\mathrm{nd}}$ are concatenated and fed in the three modules commonly used in industry mentioned above with other features, i.e.,  item features, user profiles, and recommendation contexts. 

\textcolor{black}{In addition to DFN, two other works also denoise the implicit feedback by an attention network with the help of the explicit feedback, the first of which, DUMN~\cite{RecSys2021-DUMN}, also utilizes a memory network for modeling users' long-term preferences to perform the CTR prediction task, while the second work FeedRec~\cite{WWW2022-FeedRec}, a work focusing on news recommendation, uses Transformers with shared and unshared parameters to perform user modeling.}

\textbf{NextIP.}
A dual-task learning approach towards \textcolor{black}{the item prediction} task and purchase prediction task (NextIP)~\cite{CIKM2022-NextIP} utilizes the self-attention mechanism to model \textcolor{black}{multi-behavior} sequences from a local modeling perspective \textcolor{black}{and performs the next-item recommendation task}. Unlike other methods, NextIP simultaneously treats the \textcolor{black}{multi-behavior} sequences as some sequences of behavior-specific items and a sequence of (item, behavior) pairs. Specifically, NextIP treats the \textcolor{black}{multi-behavior} sequential recommendation problem as two tasks, i.e.,  \textcolor{black}{the item prediction} task and the purchase prediction task.

In \textcolor{black}{the item prediction} task, the embeddings of \textcolor{black}{behavior-specific  and behavior-aware item sequencesare entered into the self-attention block (SAB).} 
Subsequently, NextIP proposes the target-behavior-aware context aggregator (TBCG) to fully \textcolor{black}{model} the interplay of different behaviors at different times. Specifically, TBCG takes the representations of the most recent interaction for behavior-specific subsequences as keys and values, takes the user's target behavior embedding as a query, and inputs those into the attention module and mean pooling function with the target behavior representations from the behavior-specific subsequence representations. Finally, \textcolor{black}{the item prediction} result is calculated by the inner product between the target item embedding and the representation added by the output of TBCG and the most recent interaction representation of the \textcolor{black}{behavior-aware} sequences.

In the purchase prediction task,  the user's behavior sequence embeddings are input into the behavior-aware self-attention block, masked depending on user behavior types and behavior distance. \textcolor{black}{Each} auxiliary behavior representation from the output of the behavior-aware self-attention block is treated as negative samples to \textcolor{black}{model} the user purchase preference. 

In summary, NextIP proposes a new perspective on this multi-behavior sequential recommendation problem,  by framing it as both an item prediction and a purchase prediction task. This new perspective offers a fresh outlook on the issue at hand, allowing for more accurate and efficient solutions. Moreover, NextIP considers multiple input patterns of the multi-behavior sequences and uses the self-attention network to model multi-behavior sequences with good performance. \textcolor{black}{The contrastive loss function used to train the model also contributes to recommendation performance.}

\textbf{MB-STR.}
Multi-behavior sequential Transformer recommender (MB-STR)~\cite{SIGIR2022-MB-STR}  utilizes Transformer to model \textcolor{black}{multi-behavior} sequences from both global and local modeling perspectives, \textcolor{black}{to address the next-item recommendation problem}. \textcolor{black}{MB-STR treats the multi-behavior sequence as a sequence of (item, behavior) pairs and feeds it into the multi-head self-attention network, which considers the sequential pattern and distinguishes it based on the types of behavior. Then a parameter-shared network like MMoE is used to model the behavior-specific information, denoted as a Behavior Aware Prediction (BA-Pred) module. BA-Pred includes two parts, i.e., the \textcolor{black}{parameters-shared} experts and the behavior-specific experts, where the latter are shared for the representations of the same behavior.}

In summary, MB-STR employs a range of behavior-specific parameters to represent diverse behavioral sequences at a fine-grained level. This approach enables effective modeling of the distinctiveness and interdependence among various behaviors, rendering it a robust tool for behavior modeling.
\textcolor{black}{Meanwhile, the total number of parameters in MB-STR is $O(|\mathcal{V}|d+|\mathcal{B}|d^2+n)$, and its time complexity is $O(n^2d+nd^2)$, which is moderate compared to other works.}
Moreover, unlike the positional encoding function of the classical Transformer, MB-STR is inspired by T5~\cite{raffel2020exploring} in natural language processing and uses a novel positional encoding function to model multi-behavior sequence relationships, which can better capture their positional relationships.

\textbf{FLAG.}
Feedback-aware local and global (FLAG)~\cite{TIST2022-FLAG} takes into account both user intent and preference complexity in modeling multi-behavior sequences for next-item recommendation.
It takes a behavior-agnostic sequence of items and a sequence of behaviors as input, and employs both the global and local modeling perspectives. FLAG has four \textcolor{black}{parts}, including a local preference \textcolor{black}{modeling}, a global preference \textcolor{black}{modeling}, a local intention \textcolor{black}{modeling} and a prediction module.

\textcolor{black}{
In the local preference modeling, the input matrix $X_{u}^{(0)}$, composed of the element-wise additions of the item embedding and the position embedding, is fed into the multiple stacked feedback-aware self-attention blocks (FSABs), and then obtains a user’s local preference $ \boldsymbol{z}_{t}^{\mathrm{lp}}$ at time step $t$ from the top FSAB. 
Specifically, an FSAB successively goes through a feedback-aware input layer with a mask mechanism, a self-attention layer and a feed-forward layer.}
\textcolor{black}{In the global preference modeling, the authors use a location-based attention layer to model users' global preferences $\boldsymbol{z}^{\mathrm{gp}}$.}
Given that the preferences of users, both local and global, cannot be effectively modeled through local preference modeling and global preference modeling alone,
\textcolor{black}{
a feedback-based attention layer (FAL) is proposed for local intention modeling. It receives an input matrix $\boldsymbol{O}$ that takes into account both the examination-specific and purchase-specific embedding matrices:
}
\begin{align}
    &\boldsymbol{o}_{u}^{t}=\textcolor{black}{V_{i_{u}^{t}}}+\boldsymbol{p}_{t}^{\prime}+\textcolor{black}{F_{f_{u}^{t}}}\\
    &\boldsymbol{O}=\left[\boldsymbol{o}_{u}^{1} ; \ldots ; \boldsymbol{o}_{u}^{l} ; \ldots ; \boldsymbol{o}_{u}^{T}\right]
\end{align}
where $V_{i_{u}^{t}} \in \mathbb{R}^{1 \times d}, \boldsymbol{p}_{t}^{\prime} \in \mathbb{R}^{1 \times d} \text { and } F_{f_{u}^{t}} \in \mathbb{R}^{1 \times d}$ are the item-specific embedding vector, the position-specific embedding vector and the behavior-specific embedding vector $ f_{u}^{t}$ of the item $i_{u}^{t}$ at time step $t$, respectively.
\textcolor{black}{And the next behavior $F_{f_{u}^{t+1}}$ is treated as a query vector to uncover the user's local intention in the following time step, so as to obtain the final local intention feature $\boldsymbol{z}_{t}^{l i}$}.
\textcolor{black}{Then an item similarity gating (ISG) module is proposed}
to achieve a balance between the local and global preferences with a weight factor $\lambda$, and then the obtained balanced preference representation $\boldsymbol{z}_{t}^{\mathrm{lgp}}$ and the local intention feature $\boldsymbol{z}_{t}^{\mathrm{li}}$ are element-wise added to get the final representation $\boldsymbol{z}_{t}$  of the sequence at time step $t$.

\textcolor{black}{FLAG models the user's local preference, global preference and local intention \textcolor{black}{with acceptable time complexity and space complexity}, where the multiple behaviors is utilized as a mask matrix in the local preference learning module, and as part of the input to the module through behavior embedding for better distinguishing the user’s different behaviors and consequently improve preference modeling. However, in the local intention learning module, FLAG uses the next real feedback as the query vector during training, which may have a data bias that allows the model to overfitting the historical behavioral data. Furthermore, this approach may not perform well in cold-start settings where there is little historical interaction data.}

\textcolor{black}{In summary,} Transformer, a sequence-to-sequence model, has demonstrated exceptional performance in recommender systems. Typically, Transformer captures the temporal relationship of behaviors by incorporating positional information in MBSR. Through the utilization of an attention mechanism, it is able to model relationships both within and between behaviors. \textcolor{black}{With superior parallel computing capabilities, an enhanced ability to capture long-term dependencies, and stronger interpretability, Transformer surpasses RNN and GNN in MBSR to some extent.}
\subsection{Generic-Methods-based Learning Architecture}
Since there are a lot of relevant and advanced works in a research area, it is necessary to study a generic framework that can utilize any of the previous relevant works to obtain information. A learning architecture based on a generic method that can \textcolor{black}{employ a particular designed module on a state-of-the-art model}, combined with some innovative modeling modules to \textcolor{black}{enhance}
 the performance of that model, which is a direction worth further study.

\subsubsection{Methods in MBSR}
For the MBSR problem, the most important issues to consider are how to model sequences and how to distinguish between different behaviors. As such, the use of generic-methods-based learning architectures can be chosen to improve the recommendation performance by following previous effective models of SBSR or MBR in the modeling of sequences or heterogeneous behaviors. Behavior-aware recommendation (BAR)~\cite{IS2022-BAR} is a generic framework utilized in terms of obtaining sequence representations, which we introduce below.

\begin{figure}[!htb]
	
	\begin{center}
		
		\begin{tabular}{cc}
			
			\includegraphics[width=2.5in]{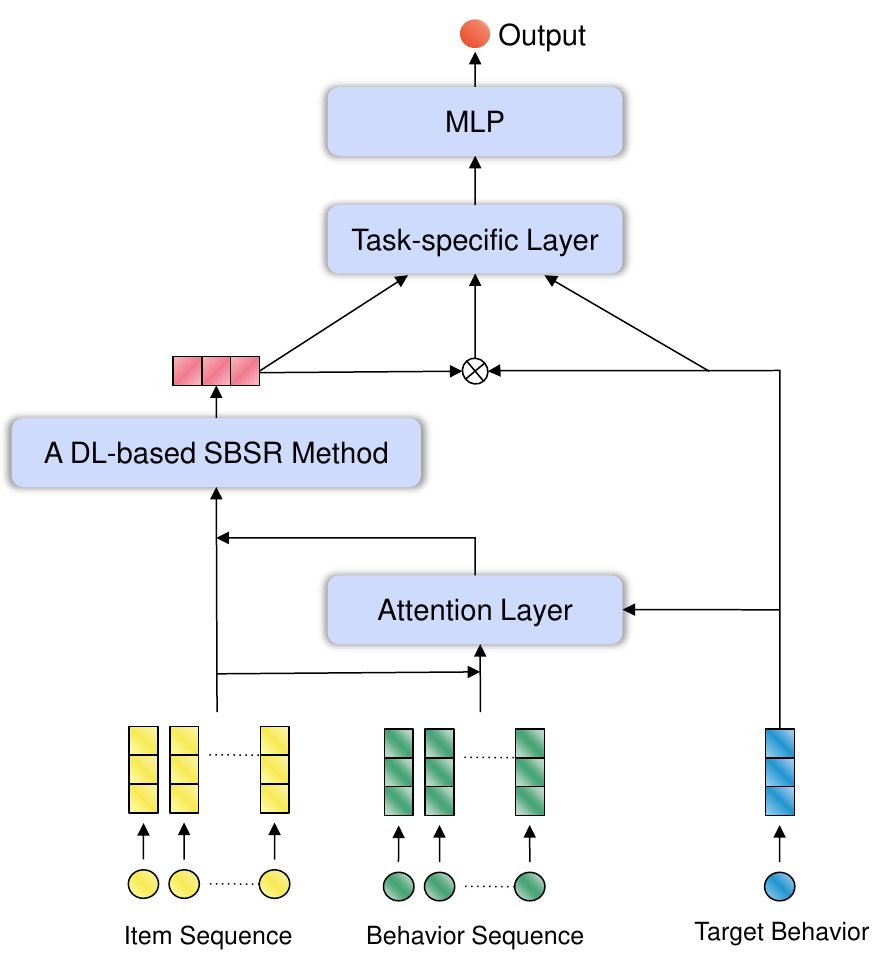}
			
		\end{tabular}
		
	\end{center}
	
	\caption{Illustration of behavior-aware recommendation (BAR).}
	
	\label{fig:BAR}
\end{figure}

\textbf{BAR.}
BAR proposes a generic learning architecture for modeling \textcolor{black}{multi-behavior} sequences from a global modeling perspective, including a behavior attention layer and a task-specific layer, as we illustrate in Figure~\ref{fig:BAR}.
In the behavior attention layer, an attention network is used to enhance the presentation of the item embedding. 
Firstly, the embedding of an item ${\ell}$ is added by the  behavior embedding $B_{b_{u}^{\ell}}$ and the position embedding $P_{\ell}$.
\textcolor{black}{Then an attention network is used to obtain the attention score $\alpha_{\ell} \in \mathbb{R}$ representing the relationship between the behavior embedding $B_{b_{u}^{\ell}}$ and the new presentation of item embedding $X_{\ell}$, and is added to the item embedding $V_{i_{u}^{\ell}}$ to learn the hidden representation at each time step:}
\begin{align}
&\boldsymbol{h}_{t-1}=\mbox{RM}\left(\left(1+\alpha_{t-L}\right) V_{i_{u}^{t-L}}, \ldots,\left(1+\alpha_{t-1}\right) V_{i_{u}^{t-1}}\right)
\end{align}
where $\mbox{RM}(\cdot)$ denotes some important components used in sequential recommendation methods, e.g., recurrent neural network and convolutional neural network. $\mbox{RM}(\cdot)$ reflects the generality of BAR, as any SBSR method like SASRec~\cite{ICDM2018-SASRec} can be utilized as a module of $\mbox{RM}(\cdot)$ to learn the potential representations of sequences.

The task-specific layer is proposed as a solution to address the challenge of \textcolor{black}{unknown} whether the behavior is the purchase or not when the model is focused on predicting the next purchased item.
\textcolor{black}{It uses an MLP to obtain the \textcolor{black}{connection} between the sequential information representation  $\boldsymbol{h}_{t-1}$ and the behavior embedding $B_{b_{u}^{\ell}}$.}

\textcolor{black}{In summary, a general framework like BAR with directly applying the modeling methods used in SBSR possesses better performance and strong generalization capability, but now there are few works aiming to enhance the performance of recommendations. Hence, it could be beneficial to investigate the generalizability of modeling behavior types and transitions or to propose a generic model that incorporates the items' knowledge graph and the social connections among users.}
\subsection{Hybrid-Methods-based Learning Architecture}
Combining multiple technologies for modeling can make use of the advantages of different technologies, and different technologies can also complement each other, leading to the improvement of modeling ability.
\textcolor{black}{The effective integration of diverse technologies within different modules is a crucial aspect to be considered when utilizing a hybrid-methods-based learning architecture.}

\subsubsection{Methods in MBSR}
MBSR needs to model the sequence and behavior types at the same time, and it also needs to consider long-term and short-term preferences, as well as local or global information, which provides opportunities for employing different technologies. In MBSR, there are some works utilizing different techniques, including MKM-SR~\cite{SIGIR2020-MKM-SR}, MBGNN~\cite{MLBDBI2021-MBGNN}, MBHT~\cite{KDD2022-MBHT}, KHGT~\cite{AAAI2021-KHGT} and TGT~\cite{TKDE2022-TGT}. We describe \textcolor{black}{some of them} in detail below, and summarize the data perspective, the modeling perspective and the characteristics of these works in Table~\ref{tbl:Hybrid-Methods}.

\begin{table*}[htbp]
\caption{Data \& modeling perspectives and features used in works based on hybrid learning architecture.} \label{tbl:Hybrid-Methods}
\begin{center}
\small
\begin{tabular}{ p{1.5cm} | p{2.6cm} | p{4cm} | p{2.5cm} |  p{5.4cm}} \hline\hline
Works & Hybrid Techniques & Data Perspective & Model Perspective & Features\\
    \hline\hline
MKM-SR~\cite{SIGIR2020-MKM-SR}
    & RNN + GNN
    & A behavior-agnostic sequence of items and a sequence of behaviors
    & Global
    & Consider the knowledge graph of the items and the attributes.\\
    \hline
MBGNN~\cite{MLBDBI2021-MBGNN}
    & RNN + GNN
    & Some behavior-specific subsequences of items
    & Local + Global
    & Consider both behavior type and direction to distinguish different node sets.\\
    \hline
KHGT~\cite{AAAI2021-KHGT}
    & Transformer + GNN
    & Some behavior-specific subsequences of items
    & Local + Global
    &  Consider item-item relation information.\\
    \hline
MBHT~\cite{KDD2022-MBHT}
    & Transformer + GNN
    & A sequence of (item, behavior) pairs
    & Local + Global
    & Model users' short-term and long-term preferences by self-attention network and graph neural network, respectively.\\
    \hline
TGT~\cite{TKDE2022-TGT}
    & Transformer + GNN
    & A sequence of (item, behavior) pairs
    & Local + Global
    & Model long-term and short-term multi-behavior sequence features separately to \textcolor{black}{model} a user's dynamic preference.\\
\hline\hline 
\end{tabular}
\end{center}
\end{table*}
\textbf{MKM-SR.}
Micro-behaviors and item knowledge into multi-task learning for session-based recommendation (MKM-SR)~\cite{SIGIR2020-MKM-SR}  utilizes the gated graph neural networks (gated GNN, GG-NNs)~\cite{wu2019session} and the gated recurrent units (GRU) to model \textcolor{black}{multi-behavior} sequences from a global modeling perspective. Here the global modeling perspective denotes that all sequences are modeled together, rather than each sequence separately. We focus on the part of MKM-SR that models user \textcolor{black}{multi-behavior} sequential information, i.e., M-SR.
It treats the \textcolor{black}{multi-behavior} sequence as a sequence of items and a sequence of behaviors modeling.
\textcolor{black}{Then M-SR utilizes GG-NNs and GRU to model the item sequence and the behavior sequence, respectively, and concatenates the output vectors to obtain the behavioral characteristics of the user.}

M-SR aggregates the embedding of the nodes by the constructed user-item graph.
Subsequently, it is fed into the GRU module to enhance the information further.
In comparison with the methods utilizing \textcolor{black}{RNN} alone, M-SR can capture the bidirectional sequence relationships. \textcolor{black}{Furthermore, M-SR's methodology of framing the item sequence and inputting it into the GGNN proficiently models the relationships among all items. Additionally, utilizing the GRU to input behavioral sequences, rather than GGNN, enables M-SR to effectively capture the user's behavioral sequential preferences.}

\textbf{KHGT.}
Knowledge-enhanced hierarchical graph Transformer network (KHGT)~\cite{AAAI2021-KHGT} also utilizes Transformer and graph neural network to model \textcolor{black}{multi-behavior} sequences from both the global and local modeling perspectives, and treats the \textcolor{black}{multi-behavior} sequence as some sequences of behavior-specific items modeling. 
For the position information of the user \textcolor{black}{multi-behavior} sequences, KHGT designs a novel encoding position function, \textcolor{black}{which takes into account the users, the items, and the behavior types.}
For the user-item graph, unlike other methods, it constructs a heterogeneous graph of all users and \textcolor{black}{interacted} items. Each edge represents a record of a user's interaction with an item under a certain behavior type. The item-item graph is constructed using the item relation information, such as the item category. 
\textcolor{black}{To extract  the transition information  about the nodes, a behavior-specific multi-head self-attention network is employed}, and \textcolor{black}{then the information of the graphs is utilized} to aggregate the neighborhood information of the learning node. Finally, the information of each node is obtained.

\textcolor{black}{KHGT is one of the few approaches to incorporate an item-to-item relationship within MBSR. This integration effectively enhances the information pertaining to each item, resulting in improved recommendation performance.} It constructs the user-item and item-item graphs, and uses Transformer to model the relationship between different behaviors. 
\textcolor{black}{The relationships within each behavior and between multiple behaviors are thoroughly} considered and are thus modeled well.

\textcolor{black}{Apart from the above two works, MBGNN~\cite{MLBDBI2021-MBGNN} leverages GRU and GNN to model the user's global and local preferences, respectively, to solve the session-based recommendation problem, where behavior transition is considered in the construction of the graph similar to other GNN-based works. Whereas, both MBHT~\cite{KDD2022-MBHT} and TGT~\cite{TKDE2022-TGT} utilize GNN and Transformer to model users' long-term and short-term preferences, respectively, where the former designs a novel self-attention mechanism inspired by Linformer~\cite{wang2020linformer}.}
\textcolor{black}{In summary, the increasing use of hybrid-methods-based learning architecture for works in MBSR suggests that combining different techniques can leverage the strengths of these techniques and play a complementary role, thus enhancing the recommendation performance. Consequently, this is a direction worthy of further research.}

\section{Future Directions}
\textcolor{black}{The multi-behavior sequential recommendation (MBSR) problem, which is more representative of real-world recommendation scenarios, has increasingly gained attention from academia and industry in recent years.}
Although some works with superior recommendation performance towards the MBSR problem have been proposed, there are still many issues worthy of further study.  \textcolor{black}{In this section, we discuss some potential future research directions} for the MBSR problem, including data, techniques, \textcolor{black}{optimization targets and trustworthiness and responsibility.}

\textit{Data}. In the field of artificial intelligence, a comprehensive understanding of data is crucial for developing models. In the case of MBSR, the complexity of the data also poses various challenges when modeling.
First of all, data sparsity has always been the focus of recommendation algorithms~\cite{guo2017DataSparsity}, and MBSR is no exception. 
However, excessive data sparsity can undermine the performance of association-based algorithms like collaborative filtering in recommender systems. Additionally, the multiple behaviors of MBSR make the pattern of data sparsity more intricate. In practical situations, such as cold-start settings, where new users or items are seldom interacted with, resolving the data sparsity issue is necessary to generate reasonable recommendations.
Secondly, it is essential to explicitly model the data imbalance in MBSR. The data suffers from a heterogeneous behavioral distribution problem similar to MBR and a sequence length problem similar to SBSR. User behavior distribution and interaction sequence lengths often differ in real-world scenarios. For instance, in shopping scenarios, users tend to make fewer purchases than examination behaviors, and \textcolor{black}{users may examine varying item quantities.}
Thirdly, there are several issues associated with data processing, including periodicity and noise. Periodicity refers to users' inclination to examine items at specific times, and noise refers to users examining items that do not align with their current preferences. While related works have focused on denoising~\cite{IJCAI2020-DFN,RecSys2021-DUMN}, there remains a significant scope for further research, particularly in terms of how to explicitly model various types of specific noise, such as interactions that align with a user's long-term preferences but not their current preferences.
As such, it is necessary to further explore how to deal with data sparsity, imbalance, periodicity and noise, etc, so as to improve the effectiveness of recommendations.

\textit{Techniques}. Technical innovation has been the approach that most works have been focused on to improve the recommendation performance, and there are several challenges to the techniques currently used for MBSR. Firstly, single types of techniques have their own limitations.
For example, Transformer can solve the problem of parallel computation that RNN is limited, but is less capable of capturing the local information than RNN due to the point-wise dot-product self-attention utilized~\cite{NeurIPS-2019-TransformerVsRNN,arXiv2022-HGN}. 
\textcolor{black}{As such, combining multiple complementary components or techniques to solve the MBSR problem is an important research direction.} 
Secondly, efficiency is an essential issue in MBSR due to the complexity of the data. It is worthwhile to investigate how to improve recommendation performance without sacrificing efficiency so as to enable real-time recommendations. 
\textcolor{black}{Thirdly, how to maintain acceptable time and space complexity when the number of behavior types increases is also a challenging issue.}
Fourthly, some works propose models that perform well on some datasets but poorly on others during training and prediction~\cite{RecSys2022-GPG4HSR,TIST2022-FLAG}. As such, it remains a challenge to improve the generalization of the models for MBSR.
In addition, there are difficulties in investigating the MBSR problem utilizing data from different domains, or data with auxiliary information such as item category information, reviews, and knowledge graphs. As interactive conversational recommender systems become more prevalent between users and platforms, future MBSR techniques may need to model multi-behavior sequential data and multiple rounds of conversational text data. In summary, there is much valuable research that can be done on the technical aspects of the MBSR problem, especially in terms of combining methods, improving efficiency, and adaptability of data diversity.

\textit{Optimization targets}. \textcolor{black}{Optimizing targets in MBSR also presents several challenges. }Currently, most works on MBSR focus on a single target, such as recommending more items that users would like to buy in a shopping scenario. However, \textcolor{black}{the diversity of user behaviors allows the possibility of optimizing multiple targets simultaneously. For example,} on the business side of the industry, there is often more than one single target to optimize~\cite{wang2019multi-target}, instead, there is a need to jointly optimize multiple targets, such as increasing the view rate and like rate of a video simultaneously. \textcolor{black}{At present, the multi-target optimization methods mainly include setting sample weight, stacking multiple models, sharing model parameters for joint training and MMoE~\cite{KDD-2018-MMoE}, etc. However, there are some shortcomings in these methods. For example, in the method of stacking multiple models, the models are independent of each other, which makes the training process prone to the situation of over-fitting, while the sharing of experts in MMoE among all tasks may bring bias to some tasks. As such,} \textcolor{black}{how to optimize multiple objectives in a rational way} is also a direction worth investigating.

\textit{LLMs}. The remarkable performance of large language models \cite{floridi2020gpt, zhao2023survey} has received great attention within the academic community. A mounting body of research is presently dedicated to expansive language models in the domain of recommendation systems\cite{wu2023survey, fan2023recommender}. Due to the large amount of textual information intrinsic to the recommendation task itself, as well as the commendable language comprehension ability and external knowledge reserve of the large model, modeling the representation of users and items with text information may hopefully supplant conventional ID-based paradigms. 
In the task of multi-behavior sequential recommendation, in addition to modeling sequence relationships, different behavior information also needs to be modeled for items. How to use a large model to effectuate a synergistic amalgamation of distinct behavioral signals with the textual profiles of users and items is a relatively new direction. Noting that items may have different relationships under different behaviors. To illustrate, if you have browsed a certain type of product, you may buy the same type of product from a different brand. But buying this type of item may not buy it from a different brand again. Therefore, it is necessary to integrate multi-behavioral information into the modeling of large models and recommendation tasks, rather than simple sequence modeling.

\textit{Trustworthiness and responsibility}. The need to build more trustworthy and responsible recommender systems has been raised when recommender systems consistently pursue higher accuracy, and are determined to recommend items to users transparently, fairly and unbiasedly. 
Explainability and security are two main aspects of the trustworthiness of recommender systems~\cite{wang2022trustworthy}, which also require further attention in MBSR. Firstly, in terms of explainability, \textcolor{black}{the complexity of the behaviors in MBSR makes the deep learning model less explainable.} Attention is a common approach applied to MBSR to improve the explanation of deep learning models~\cite{WSDM2018-RIB,IJCAI2020-DFN,KDD2022-MBHT,IS2022-BAR}.
Secondly, in terms of security, the issue of privacy protection is becoming increasingly important to the state and to the public. Recommender systems need to avoid the problem of user information leakage when designing a model, including the risk of information leakage between users, between platforms, and between both users and platforms. 
For the MBSR problem, the user's behaviors are considered private.
Such private-sensitive data can only be observed on the user’s own client and cannot be uploaded to the cloud, thereby complicating the modeling process. To address this challenge, the federated recommendation,
\textcolor{black}{one of the most effective and popular approaches addressed the privacy protection problem in recommender systems, is first proposed by Google in 2016~\cite{google2017federated-leanring}.}
Not much work has been done to consider privacy and security in MBSR~\cite{WWW2021-DeepRec}, and the use of federated learning~\cite{TIST2019FederatedLearning} to secure privacy is an interesting direction. 
As such, it is important to build trustworthy and responsible recommender systems with higher explainability under the requirement of privacy protection, so that users can be fairly recommended the items they are interested in.

\section{Conclusions}
MBSR combines SBSR and MBR, requiring to model both sequential information and heterogeneous behaviors, which provides some challenges while allowing for some optimization in recommendation performance. MBSR is closer to the range of user feedback that occurs in real-life scenarios. The increasing amount of works for MBSR in academia and industry, although much fewer than those for SBSR, indicates the importance of MBSR \textcolor{black}{in recommender systems}. In this paper, we first introduce the MBSR problem in detail, followed by a classification of related works, encompassing neighborhood-based methods, matrix factorization-based methods, and deep learning-based methods.

Due to the complexity of the MBSR problem, lots of works \textcolor{black}{use deep learning-based methods, including RNN, GNN, and Transformer, or their generic and hybrid architectures.} For each of these learning architectures, we present their general form before transitioning to how to apply the learning architecture to the MBSR problem based on previous works. For each work, we introduce it by technology, data and modeling perspectives, and discuss its strengths and weaknesses.

Through a detailed discussion of the works that have been done so far, we find that MBSR still faces many challenges. In response to these challenges, we suggest five possible future directions, including data, techniques, security, optimization targets and explanation, which we hope will give the readers some guidance on how to solve the MBSR problem better.

\section*{Acknowledgements}
We thank the support of National Natural Science Foundation of China Nos. 62172283 and 62272315, and Mr. Jinwei Luo for helpful discussions.

\bibliographystyle{IEEEtran}

\bibliography{paper}

\end{document}